\documentclass[10pt,journal,compsoc]{IEEEtran}
\usepackage{multirow}    
\usepackage{pifont}      
\usepackage{color}       
\usepackage{xspace}      
\usepackage[keeplastbox]{flushend}     
\usepackage{graphicx}
\usepackage{caption}
\usepackage{subcaption}
\usepackage{url}
\usepackage{rotating}
\usepackage{paralist}

\usepackage{enumitem}
\usepackage{soul}

\usepackage{algorithm}  
\usepackage{algorithmicx}  
\usepackage{algpseudocode}  
\usepackage{amsmath} 
\algnewcommand{\LineComment}[1]{\State \(\triangleright\) #1}
\usepackage{color}
\usepackage{tikz}
\usepackage{xcolor}

\begin{document}
\newcommand{\circled}[1]{\tikz[baseline=(myanchor.base)] \node[circle,fill=.,inner sep=1pt] (myanchor) {\color{-.}\bfseries\footnotesize #1};}
\newcommand{\mypara}[1]{\vspace{2pt}\noindent\textbf{{#1: }}}
\newcommand{\eat}[1]{}  
\newcommand{\name}{Attack\xspace}
\newcommand{\lowerambush}{memory ambush\xspace}
\newcommand{\upperambush}{Memory Ambush\xspace}
\newcommand{\authcomment}[3]{\textcolor{#3}{#1 says: #2}}
\newcommand{\zhi}[1]{\authcomment{zhi}{#1}{blue}}
\newcommand{\yueqiang}[1]{\authcomment{yueqiang}{#1}{red}}
\newcommand{\henry}[1]{\authcomment{henry}{#1}{green}}
\renewcommand{\baselinestretch}{0.96}   

\definecolor{R}{RGB}{0,0,150}

\title{CATTmew: Defeating Software-only \\ Physical Kernel Isolation}
%
%
%

\author{Yueqiang~Cheng$^*$,
        Zhi~Zhang$^*$,
        Surya~Nepal,
        Zhi~Wang 

\IEEEcompsocitemizethanks{
\IEEEcompsocthanksitem Yueqiang Cheng and Zhi Zhang are joint first authors.\protect
\IEEEcompsocthanksitem Yueqiang Cheng is with Baidu XLab, America.\protect\\
E-mail: chengyueqiang@baidu.com 

\IEEEcompsocthanksitem Zhi Zhang is with the Data61, CSIRO, Australia and the University of New South Wales, Australia.\protect\\
E-mail: zhi.zhang@data61.csiro.au 

\IEEEcompsocthanksitem Surya Nepal is with the Data61, CSIRO, Australia.\protect\\
E-mail: surya.nepal@data61.csiro.au
\IEEEcompsocthanksitem Zhi Wang is with the Department of Computer Science, Florida State University, America.\protect\\
E-mail: zwang@cs.fsu.edu}}
\IEEEtitleabstractindextext{%
\begin{abstract}
All the state-of-the-art rowhammer attacks can break the MMU-enforced inter-domain isolation because the physical memory owned by each domain is adjacent to each other. 
To mitigate these attacks, physical domain isolation, introduced by CATT~\cite{brasser17can}, physically separates each domain by
dividing the physical memory into multiple partitions and keeping each partition occupied by only one domain. CATT implemented physical kernel isolation as the first generic and practical software-only defense to protect kernel from being rowhammered as kernel is one of the most appealing targets.


In this paper, we develop a novel exploit that could effectively defeat the physical kernel isolation and gain both \emph{root} and \emph{kernel} privileges.
Our exploit can work without exhausting the page cache or the system memory, or relying on the information of the virtual-to-physical address mapping. 
The exploit is motivated by our key observation that the modern OSes have \emph{double-owned} kernel buffers (e.g., video buffers and SCSI Generic buffers) owned concurrently by the kernel and user domains. The existence of such buffers invalidates the physical kernel isolation and makes the rowhammer-based attack possible again.
Existing conspicuous rowhammer attacks achieving the root/kernel privilege escalation exhaust the page cache or even the whole system memory. Instead, we propose a new technique, named \emph{\lowerambush}. It is able to place the hammerable double-owned kernel buffers physically adjacent to the target objects (e.g., page tables) with only a small amount of memory. As a result, our exploit is stealthier and has fewer memory footprints.
We also replace the inefficient rowhammer algorithm that blindly picks up addresses to hammer with an efficient one. Our algorithm selects suitable addresses based on an existing timing channel~\cite{moscibroda2007memory}.
We implement our exploit on the Linux kernel version 4.10.0.
Our experiment results indicate that a successful attack could be done within $1$ minute. The occupied memory is as low as $88$\emph{MB}.
\end{abstract}

}

\maketitle


%
\IEEEpeerreviewmaketitle

\section{Introduction}\label{sec:intro}
A memory management unit (MMU) is an essential component of the CPU. It plays a critical role in enforcing isolation in the operating system (OS). For example, the kernel relies on the MMU to mediate all memory accesses from user processes in order to prevent them from modifying the kernel or accessing its sensitive information. Any unauthorized access will be stopped with a hardware exception. Without the strict kernel-user isolation, the whole system can be easily compromised by a malicious user process, such as a browser~\cite{gruss2016rowhammer}. The MMU is also used widely in other forms of isolation, such as intra-process isolation (e.g., sandbox) and inter-virtual machine (VM) isolation. Therefore, the MMU and its key data structure, page tables, are critical to the security of the whole system. However, the recent rowhammer attacks have posed a serious challenge to the status quo. 

\mypara{Rowhammer attacks}
Dynamic Random Access Memory (DRAM), the main memory unit of a computer system, is often organized into rows. Kim et al.~\cite{kim2014flipping} discovered that intensive reading from the same addresses in two DRAM rows (i.e., aggressor rows) can cause bit flips in an adjacent row (i.e., a victim row). By placing the key data structures, such as page tables, in a victim row, an attacker can corrupt these data structures by ``hammering'' the aggressor rows. Such attacks are collectively called ``rowhammer'' attacks. Rowhammer attacks can stealthily break the MMU-enforced isolation because they do not need to access the victim row at all, and they do not rely on any design or implementation flaws in the isolation mechanisms. Rowhammer attacks have been demonstrated to break all popular forms of isolation: 

\begin{itemize}[itemsep=1ex,leftmargin=0.4cm]
		\item \emph{Intra-process isolation:} this isolation separates the untrusted code from the trusted code within a single process. The untrusted code (e.g., JavaScript) can break the isolation and gain a higher privilege (e.g., the browser's privilege)  by exploiting rowhammer vulnerabilities~\cite{seaborn2015exploiting, qiao2016new, gruss2016rowhammer, bosman2016dedup, frigo2018grand, tatar2018throwhammer}.
		
		\item \emph{Inter-process isolation:} this is essentially the process isolation enforced by the OS kernel. Rowhammer attacks can break the process boundary to steal private information (e.g., encryption keys)~\cite{bhattacharya2016curious}  and break code integrity (e.g., to gain root privilege by flipping opcodes of a \emph{setuid} process)~\cite{gruss2017another}. 
		
     \item \emph{Kernel-user isolation:} this isolation protects the kernel from user processes. Rowhammer attacks have been shown to break this isolation on both x86 and ARM architectures~\cite{seaborn2015exploiting, van2016drammer}.
    \item \emph{Inter-virtual machine (VM) isolation:} this isolation protects one VM from another. Inter-VM isolation is especially important in the cloud environment where VMs from different customers can co-exist on the same physical machine. By leveraging rowhammer attacks, a malicious VM can break the inter-VM isolation and tamper with the victim VM's code or data (e.g.,  OpenSSH public key and page tables)~\cite{razavi2016flip, xiao2016one}. 
    \item \emph{Hypervisor-guest isolation:} this isolation protects the hypervisor from its guest VMs. Rowhammer attacks have been demonstrated on the paravirtualized (PV) platform against this isolation~\cite{xiao2016one}.
\end{itemize}
As such, rowhammer attacks pose a serious threat to the security of computer systems.

\begin{table*}
\footnotesize
\centering
\begin{tabular}{ccccc}
\hline
\multirow{2}{*}{\textbf{Key Technique of Rowhammer Exploits}} & \multirow{2}{*}{\textbf{Memory Usage}} & \multirow{2}{*}{\textbf{Page Cache Usage}} & \multirow{2}{*}{\textbf{Severe Privilege Escalation}} & \multirow{2}{*}{\textbf{Defeat CATT~\cite{brasser17can}}}\\  
 &  &  &  &  \\ \hline
\multirow{2}{*}{Memory Spray~\cite{seaborn2015exploiting}} & \multirow{2}{*}{Exhausted} & \multirow{2}{*}{Low} & \multirow{2}{*}{Root and Kernel Privileges} & \multirow{2}{*}{${\times}$} \\
 &  &  &  &  \\ \hline
\multirow{2}{*}{Memory Groom~\cite{van2016drammer}} & \multirow{2}{*}{Exhausted} & \multirow{2}{*}{Low} & \multirow{2}{*}{Root and Kernel Privileges} & \multirow{2}{*}{${\times}$} \\
 &  &  &  &  \\ \hline
\multirow{2}{*}{Memory Waylay~\cite{gruss2017another}} & \multirow{2}{*}{Low} & \multirow{2}{*}{Exhausted} & \multirow{2}{*}{Root Privilege} & \multirow{2}{*}{${\surd}$${^\star}$} \\
 &  &  &  &  \\ \hline
\multirow{2}{*}{Throwhammer~\cite{tatar2018throwhammer}} & \multirow{2}{*}{{Exhausted}} & \multirow{2}{*}{Low} & \multirow{2}{*}{{Root Privilege}} & \multirow{2}{*}{{${\surd}$}${^\star}$}  \\
 &  &  &  &  \\ \hline
\multirow{2}{*}{\textbf{\upperambush}} & \multirow{2}{*}{\textbf{Low}} & \multirow{2}{*}{\textbf{Low}} & \multirow{2}{*}{\textbf{Root and Kernel Privileges}} & \multirow{2}{*}{\textbf{${\surd}$}} \\
 &  &  &  &  \\ \hline
\end{tabular}
\caption{A comparison of rowhammer attacks. ${\surd}$${^\star}$ means that the attacks exploit a domain-granularity issue of CATT and fixing the issue is intuitive. The \emph{\lowerambush} technique exploits a memory-ownership issue of CATT to help an unprivileged process gain both root and kernel privileges with low memory. Fixing this issue is challenging (see more details about the two issues in Section~\ref{sec:catt}).}

\label{tab:cmpexploits}
\end{table*}

\eat{
(the system is always under the memory pressure) page-table spraying sprays the memory with page tables, hoping that at least one of them lands on a physical memory page vulnerable to Rowhammer. The next step is then to flip a bit in the vulnerable physical memory page, so that the victim page table points to an arbitrary physical memory location. Given the sprayed physical memory layout, such location should probabilistically contain one of the attacker-controlled page table pages, because when the victim page table is corrupted as intended, they cannot reliably predict
the outcome of such an operation, this may cause the Rowhammer-induced bit flip to map an unrelated memory area.

(the system is under the memory pressure for a limited period) Memory Groom~\cite{van2016drammer} surgically place an attacker-controlled page table onto the vulnerable row. Since there exists a PTE in this page table flipped as carefully crafted, such PTE will point to the page table itself when a physical address bit of the PTE is flipped.
}

\mypara{Rowhammer defenses}
both hardware~\cite{DDR4,lpDDR4} and software~\cite{shutemovpagemap,pagededuplication,seaborn2015exploiting, gruss2016rowhammer, van2016drammer, brasser17can, tatar2018throwhammer, aweke2016anvil} based defenses have been proposed against rowhammer attacks. Effective hardware defenses usually require replacing the DRAM modules. As such, they are costly and cannot be applied to some legacy systems. Firmware-based solutions to double the DRAM refresh rate alleviate the problem but have been proven to be insufficient~\cite{aweke2016anvil}. 

Among software-based defenses, physical domain isolation, introduced by CATT~\cite{brasser17can} is the first generic mitigation concept against rowhammer attacks~\footnote{ANVIL~\cite{aweke2016anvil} is the first software-based rowhammer detection approach but it suffers from false positives and high overhead.}. Based on the observation that rowhammer attacks essentially require attacker-controlled memory to be physically adjacent to the privileged memory (e.g., page tables), the CATT concept aims at physically separating the memory of different domains. Specifically, it divides the physical memory into multiple partitions and further ensures that partitions are separated by at least one unused DRAM row and each partition is only owned by a single domain. 
For example, the heap in the user space will be allocated from the user partition, and  page tables are allocated from the kernel partition. 
By doing so, it can confine bit-flips induced by one domain to its own partition and thus prevent rowhammer attacks from affecting other domains.
Although current CATT implementation only enforces the domain separation between the kernel and the user spaces (i.e., physical kernel isolation), its concept can theoretically be applied to multiple domains (e.g., regular and privileged processes, multiple VMs, the hypervisor and guests), thus mitigating all the previous rowhammer attacks~\cite{gruss2017another,seaborn2015exploiting, qiao2016new, aweke2016anvil, van2016drammer,razavi2016flip,xiao2016one,frigo2018grand,tatar2018throwhammer,lipp2018nethammer}. 

\mypara{Our contributions}
\eat{
the principle of CATT is sound, but its security invariants do not always hold.
Firstly, CATT presents a domain-granularity issue in the domain separation. As CATT currently implements the user and kernel isolation, multiple processes within the user domain have to share the same partition. Although bit flips are confined to a single domain, an attacker can still target critical objects in the same process or other privileged processes of the same domain and gain the root privilege (i.e., Throwhammer~\cite{tatar2018throwhammer} and Memory Waylay~\cite{gruss2017another} shown in Table~\ref{tab:cmpexploits}).
CATT can fix the issue by further partitioning the user domain. Actually, the authors of Throwhammer have appiled CATT into their effective defense, i.e., placing attacker-hammerable buffers into an isolated domain. 
CATT can defeat Memory Waylay~\cite{gruss2017another} in a similar way (see more in Section~\ref{sec:catt}).
Also note that such exploits still cannot break the user and kernel domain isolation to directly gain the kernel privilege. 
In certain secure environments, such as containers~\cite{soltesz2007container} or high-secure systems~\cite{highsecure} with features of {\tt kernel.modules\_disabled}, {\tt kexec\_load\_disabled} and {\tt user namespace}, it is not trivial for a malicious root user to gain the kernel privilege.
}
To the best of our knowledge, we are the first to identify a \emph{memory-ownership} issue of the physical kernel isolation, that is, a block of kernel memory is initially allocated for the kernel but later mapped into the user space, allowing the user process to access the kernel memory and thus avoiding additional data copy from the user to kernel and vice versa. 
This kind of change in the memory ownership renders the physical kernel isolation ineffective, leaving the kernel still hammerable. By analyzing the Linux kernel source code, we have identified a number of such cases (more details are in Section~\ref{sec:hammerbuffer}). For brevity, we call such vulnerable memory the \emph{double-owned memory}. 
For the CATT concept itself, the physical domain isolation is also not secure if its deployment in practice does not carefully consider the performance optimization in modern OSes and thus could have a similar memory-ownership issue. 
In the rest of the paper, we discuss how to break the physical kernel isolation and further compromise the kernel, and thus use the physical kernel isolation and CATT interchangeably.



Although the aforementioned double-owned memory potentially allows a malicious process to hammer the kernel, it is still challenging to stealthily launch the rowhammer exploit. 
\emph{First}, the double-owned memory is often associated with device drivers. This limits the operations that can be performed on that memory. For example, some device drivers limit the number of the memory buffers that can be mapped to the user space. Our exploit thus needs to take these constraints into consideration. 

\emph{Second}, to position the attacker-controlled memory next to security critical objects, existing rowhammer attacks, shown in Table~\ref{tab:cmpexploits}, require exhausting either the page cache or the system memory~\cite{rowhammerjs,gruss2017another,seaborn2015exploiting, qiao2016new,aweke2016anvil,van2016drammer,xiao2016one,tatar2018throwhammer} to gain the root/kernel privilege, as summarized by Gruss et al.~\cite{gruss2017another}. Such anomaly could easily be detected by an attentive system administrator. 
{To address that, we propose a novel technique called \emph{\lowerambush} that is able to stealthily achieve the expected position with a small amount of memory (e.g., $88$\emph{MB}). Table~\ref{tab:cmpexploits} shows a comparison of our technique to other existing rowhammer exploits.}

\emph{Last}, existing single-sided rowhammer attacks~\cite{seaborn2015exploiting} cannot be simply adopted by us because they require costly random address selections but we are limited by the choice of the double-owned memory. Meanwhile, double-sided rowhammer attacks require the now-inaccessible address mapping information~\cite{seaborn2015exploiting, van2016drammer}. 
To solve this problem, we leverage the timing channel~\cite{moscibroda2007memory} to selectively pick addresses that are most likely in the same DRAM bank. This technique avoids the need to access virtual-to-physical address mapping without losing efficiency.

{To demonstrate the feasibility of our technique, we have implemented two proof-of-concept attacks against the physical kernel isolation on the Linux operating system. Our exploit uses the video and the SCSI Generic (sg) buffers as the double-owned memory and targets page tables, the critical data structure for the MMU-based isolation.}
We then use our \lowerambush technique to stealthily place either the video buffers or the sg buffers around page-table pages, by exploiting the intrinsic design of Linux's buddy allocator and {\tt mmap} syscall. After positioning the buffers next to the page-table pages, we hammer the buffer, which might flip certain bits in the page-table pages. We repeat the process until a page-table page is found to be writable. This essentially allows the attacker to read and write all system memory (i.e., kernel privilege). We also demonstrate how to gain the root privilege by changing the \emph{uid} of the current process to {\tt 0}. Our exploit can be launched by an unprivileged user process without exhausting the page cache or the system memory or relying on the virtual-to-physical
address mapping information.
{Our experiments show that the exploit can succeed within {\tt 1} minute to gain the kernel privilege and the required memory can be as low as $88$\emph{MB} with a success rate of $6$\%.}
To defend against our exploit, we have discussed possible improvements to the physical kernel isolation.



The main contributions of this paper are threefold:    
\begin{itemize}[itemsep=1ex,leftmargin=0.4cm]
    \item We identify the \emph{memory-ownership} issue of the physical kernel isolation~\cite{brasser17can}, the first practical software-based defense against rowhammer attacks and empirically demonstrate a working exploit against it.
    \item We present a novel rowhammer exploit that allows an unprivileged user process to gain the root and kernel privileges. We also discuss possible countermeasures against our exploit.  
    \item Our exploit proposes a new \lowerambush technique and a timing channel to make itself stealthy and efficient without relying on the virtual-to-physical address mapping information.

     
\end{itemize}

The rest of the paper is structured as follows.
In Section~\ref{sec:bkgd}, we briefly introduce the background information. 
In Section~\ref{sec:overview}, we present the general idea of our exploit in detail.
Section~\ref{sec:impl} demonstrates the exploit and evaluates it. In Section~\ref{sec:mitigation}, 
Section~\ref{sec:dis} and Section~\ref{sec:related}, we propose possible improvements to the physical kernel isolation against our exploit, discuss possible limitations, and summarize the related work, respectively. We conclude this paper in Section~\ref{sec:conclusion}.

\section{Background}\label{sec:bkgd}
In this section, we first describe the memory organization as it is critical to understand rowhammer attacks. We then summarize the existing rowhammer techniques.
 
\subsection{Memory Organization}
\begin{figure}
\centering
\includegraphics[width=0.85\columnwidth]{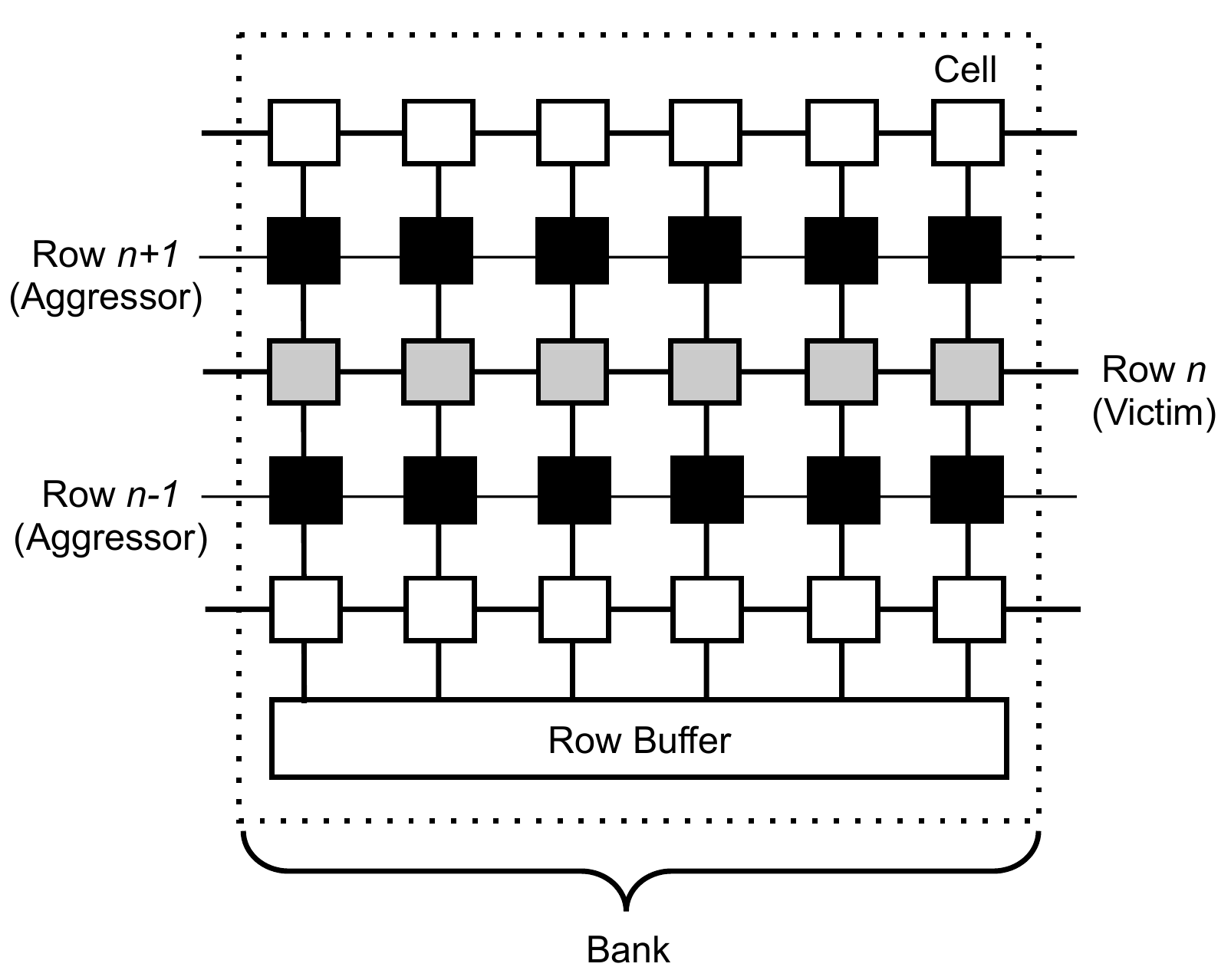}
\caption{A typical bank structure layout. In double-sided rowhammering, repeatedly accessing aggressor rows of $n+1$ and $n-1$ will trigger bit flips in victim row $n$.} 
\label{fig:bank_org}
\end{figure}

Main memory of most modern computers uses the dynamic random-access memory technology, or DRAM. Data in DRAM require periodical refresh (i.e., rewrite) to keep their value. Memory modules are usually produced in the form of dual inline memory module, or DIMM, where both sides of the memory module have separate electrical contacts for memory chips. Each memory module is directly connected to the CPU's memory controller through one of the two channels. Logically, each memory module consists of two ranks, corresponding to its two sides, and each rank consists of multiple banks. A bank is further structured as arrays of memory cells with rows and columns. For example, our test machine has a Sandy Bridge-based Core i7 CPU with  two 4GB DDR3 DIMM modules. Each module has two ranks (2GB each), and  each rank is vertically partitioned into 8 banks, which in turn consists of 32K rows of memory (8KB each). Fig.~\ref{fig:bank_org} shows the structure of a bank. Note that a typical page table in the x86-64 architecture is 4KB. 

Every cell of a bank stores one bit of data whose value depends on whether the cell is electrically charged or not. A row is the basic unit for memory access. Each access to a bank ``opens'' a row by transferring the data in all the cells of this row to the bank's row buffer. This operation discharges all the cells of the row. To prevent data loss, the row buffer is then copied back into the cells, thus recharging the cells. Consecutive access to the same row will be fulfilled by the row buffer,  while accessing another row replaces the content of the row buffer. 

\subsection{Rowhammer Overview}

\mypara{Rowhammer bugs}
Kim et al.~\cite{kim2014flipping} discovered that current DRAMs are vulnerable to disturbance errors induced by charge leakage. In particular,  their experiments have shown that frequently opening the same row  (i.e., hammering the row) can cause sufficient disturbance to a neighboring row and flip its bits without even accessing the neighboring row. Because the row buffer acts as a cache, another row in the same bank is accessed to replace the row buffer after each hammering so that the next hammering will re-open the hammered row, leading to bit flips of its neighboring row. 

\mypara{Rowhammer methods} generally speaking, there are three methods to hammer a vulnerable DRAM, classified by their memory access patterns: 

\emph{Double-sided hammering:} in this method, two immediately adjacent rows of the victim row are hammered simultaneously, as shown in Fig.~\ref{fig:bank_org}. These two adjacent rows are called the aggressor rows. They are repeatedly accessed by turn, leading to quick charges and discharges of these two rows. If the memory module is vulnerable to the rowhammer bug, this may cause some cells in the victim row to leak charge and lose their data. 

Because the aggressor rows and the victim row must lie in the same bank, this method requires at least a partial knowledge of the virtual-to-physical address mapping and the mapping between  physical addresses and the DRAM layout. {The Linux kernel originally allowed any user process to access its address mapping through the \emph{pagemap} interface. But it only allows the root user to access the interface since version 4.0~\cite{shutemovpagemap}.} Another option is to use the huge page, which allows the process to allocate a large block of continuous physical memory (2MB or 1GB). It is very likely to find two candidate aggressor rows in a huge page. However, huge pages may not be available if the kernel has the feature disabled or the memory is severely fragmented. Meanwhile, the mapping between physical addresses and the DRAM layout (i.e., DIMMs, rank, banks, and rows) can be either obtained from the processor's architectural manual or through reverse-engineering~\cite{xiao2016one,pessl2016drama}.

\emph{Single-sided hammering:}
double-sided hammering requires knowledge of the virtual-to-physical address mapping that is sometimes difficult to obtain. To this end, Seaborn et al.~\cite{seaborn2015exploiting} proposed the single-sided hammering. The main idea is to randomly pick multiple addresses and just hammer them. The probability of an aggressor row adjacent to the victim row is decided by the total number of rows.  It can be significantly improved if many aggressor rows are hammered at the same time. However, without precisely positioning of aggressor rows, this method usually induces fewer bit flips than double-sided hammering. 

\emph{One-location hammering:}
Similar to single-sided hammering, one-location hammering is also oblivious to virtual-to-physical address mappings. {But unlike single-sided hammering that hammers multiple addresses, one-location hammering ~\cite{gruss2017another} randomly selects a single address for hammering. It exploits the fact that advanced DRAM controllers employ a more sophisticated policy to optimize performance, preemptively closing accessed rows earlier than necessary. Consequently, one-location hammering does not induce row conflicts to clear the row buffer. Instead, it only has to re-open the selected row repeatedly and in the meantime, the memory controller automatically closes the row, thus inducing bit flips in a neighboring row.}

\mypara{Key requirements} there are three key requirements for exploiting rowhammer bugs: 

\emph{First}, modern CPUs employ multiple levels of caches to effectively reduce the memory access time. If data is present in the CPU cache, accessing it will be fulfilled by the cache and never reach the physical memory. As such, the CPU cache must be flushed in order to hammer aggressor rows. Even though CPU caches are mostly transparent to the user programs, they can be explicitly invalidated by instructions such as \texttt{clflush} on x86. In addition, conflicts in the cache can evict data from the cache since CPU caches are much smaller than the main memory. Therefore, to evict aggressor rows  from the cache, we can use a crafted access pattern to cause cache conflicts with the aggressor rows. Subsequent access to them will be fetched directly from the memory. 
Alternatively, we can resort to uncached memory (e.g., DMA-based buffers), since the access will not be absorbed by the CPU caches.  

\emph{Second}, the row buffer must be cleared between consecutive hammering of an aggressor row. Both double-sided and single-sided hammering explicitly perform alternate access to two or more rows within the same bank to clear the row buffer. {One-location hammering itself accesses only one row repeatedly, which lures the DRAM controller to clear the row buffer.}

\emph{Third}, for rowhammer attacks to succeed, the attacker-controlled aggressor rows must be positioned adjacent to the victim row and the victim row must contain the sensitive data (e.g., page tables) we target. Usually, the attacker does not have direct control of the (physical) memory allocation. To address that, a probabilistic approach is usually adopted on the x86 architectures. Specifically, the attacker allocates a large number of potential aggressor rows and induces the kernel to create many copies of the target objects. This strategy is very similar to the heap spray attack in that by spraying the memory with potential aggressor and victim rows, the probability of the correct positioning is high. Page tables are often targeted as the victim row because they control the system memory mapping and it is relatively easy to create many page-table pages (by allocating and using a large block of memory). An attacker-controlled page table essentially allows him to read/write/execute all the memory in the system. 

\subsection{CATT Overview and Key Observations}\label{sec:catt}

Since the kernel is the most appealing target, CATT focuses on protecting the kernel from user processes (the kernel-user isolation). As previously mentioned, {rowhammer attacks must correctly position the aggressor rows and the victim row and ensure that the victim row contains sensitive data.} CATT aims at breaking this requirement by physically separating the kernel and user memory. Specifically, it partitions each bank into a kernel part and a user part. These two parts are separated by at lease one unused row. When physical memory is allocated, CATT allocates it from either the kernel part or the user part according to the intended use of the memory. 
For example, the user heap and stacks are allocated from the user part and page tables are allocated from the kernel part. By separating the physical memory of the kernel and the user space, CATT guarantees that bit flips caused by rowhammer attacks are confined strictly into its own memory partition, thus protecting the kernel from rowhammer attacks by malicious user processes. 

{This design, however, has two potential weaknesses: 
The first is a domain-granularity issue caused by its lack of isolation in the user domain. First, even though CATT can be extended to support multiple domains, its current implementation does not support a fine-grained user domain. As such, this issue has been exploited in recently emerged rowhammer attacks~\cite{gruss2017another,tatar2018throwhammer}. 
Specifically, Throwhammer~\cite{tatar2018throwhammer} manipulates DMA buffers to compromise critical objects within a target user-process and thus breaks the intra-process isolation.  
Gruss et al.~\cite{gruss2017another} target a \emph{setuid} process that shares the user domain with the attacker process and they successfully break the inter-process isolation. Although both exploits can gain the root privilege, they cannot directly break the CATT-enforced user-and-kernel domain isolation.
In certain security scenarios, it is tough for them to further gain the kernel privilege. For security-sensitive kernel releases~\cite{highsecure}, the \texttt{kernel.modules\_disabled} is set to 1 to disallow root users from loading new kernel modules. On top of that, the \texttt{kexec\_load\_disabled} defaults to 1, disallowing the kernel memory from being altered. For virtualized systems such as containers~\cite{soltesz2007container}, a root user within a container has the same capability as a regular user in the host OS, since the \texttt{user namespace} is properly configured. As such, the host OS is well secured.}      

Intuitively, CATT can solve this issue with limited fine-grained domains. {For the intra-process isolation, one domain has the DMA buffers while the other one contains the rest of the part.} Actually, this solution has been implemented by Tatar et al.~\cite{tatar2018throwhammer} to effectively mitigate the Throwhammer. For the inter-process isolation, one domain has the regular processes while the other one contains the privileged processes.
 
{The second weakness in CATT is introduced by its static view of the memory ownership.} CATT allocates physical pages according to whether the memory is intended for the kernel or the user space. This is incompatible with the modern operating systems where the ownership of the memory is rather dynamic. For instance, some memory can be used by the device driver to send or receive data from the device and then be mapped in the user space to avoid extra copying of the data. Such memory cannot be allocated from the user partition; otherwise a crafted rowhammer exploit may badly affect the operation of the device. In the worst case, the exploit can gain the kernel privilege when the memory contains control data for the DMA operations (e.g., in the network packet transmission mechanism, the ring buffer stores {\tt transmit descriptors} that point to the packets to transmit). {On the other hand, allocating memory from the kernel partition would allow a malicious user process to hammer the kernel memory as soon as the memory is mapped to the user space.} This creates a dilemma for CATT that cannot be solved under its current design~\footnote{One naive workaround is to disable \texttt{mmap} by making an extra copy of the data from/into a user buffer when the data crosses the kernel boundary. However, this would lead to high performance overhead.}.

\eat{
In order to precisely identify which domain is issuing the memory request, CATT needs information about the context and particularly it collects the call-site information of the memory request. We contacted the CATT's authors and they confirmed that the information is from the \emph{Get-Free-Page} ($GFP$) flag. GFP flag is a Linux-specific feature, which indicates whether a calling process demanding pages is executing on behalf of a user process. And any request with one of the \emph{GFP\_USER}, \emph{GFP\_HIGHUSER} and \emph{GFP\_HIGHUSER\_MOVABLE} flags included is intended for serving the user domain, otherwise it is from the kernel domain.  
}



\section{\name Overview}\label{sec:overview}
Our primary goal is to evaluate the security of the CATT-protected kernel under our rowhammer exploit that leverages the double-owned buffers. 
In this section, we firstly present the threat model and assumptions, then identify the main challenges and introduce new techniques to overcome them. {In the next section, we present two working proof-of-concept attacks that employ the proposed techniques.} 

\subsection{Threat Model and Assumptions}
Our threat model is a little bit different from that of other rowhammer attacks~\cite{xiao2016one,razavi2016flip,qiao2016new,bosman2016dedup,gruss2016rowhammer,seaborn2015exploiting}. Specifically, 
\begin{itemize}[itemsep=1ex,leftmargin=0.4cm]
    \item The kernel is considered to be secure against software-only attacks. In other words, our exploit does not rely on any software vulnerabilities. Even though this assumption is generally not possible, we focus on the study of the rowhammer defense and attack. 
    \item The kernel is protected by CATT~\cite{brasser17can}. That is, the kernel and the user memory are allocated from physically separated partitions, and bit flips caused by rowhammer attacks are confined to their related partition.
    \item Unlike other rowhammer attacks, the attacker has no knowledge about the kernel memory locations that are bit-flippable, since CATT protects the kernel partition from being scanned.
    \item The attacker controls an unprivileged user process that has no special privileges such as accessing  {\tt pagemap}. That is, the attacker cannot obtain the virtual-to-physical address mapping. 
    \item The installed memory modules are susceptible to rowhammer-induced bit flips. Pessl et al.~\cite{pessl2016drama} report that many mainstream DRAM manufacturers have vulnerable DRAM modules,  including both DDR3 and DDR4 memory.
\end{itemize}




\subsection{Key Steps and Main Challenges}
CATT employs a static kernel/user memory partition to protect the kernel from rowhammer attacks by a malicious user process.  This implies that a physical page can only be owned by a single domain. {However, modern OS kernels often have double-owned memory that are shared between the kernel and user processes, such as video buffers and SCSI generic (sg) buffers.} If the double-owned buffer is allocated from the kernel partition, it would allow a malicious user process to hammer the kernel. 


To successfully launch a rowhammer exploit, the following five steps are necessary:  
\circled{1} identify the double-owned buffers that can be hammered;
\circled{2} stealthily position the hammerable buffers and victim kernel objects next to each other; 
\circled{3} efficiently hammer the buffer without the virtual-to-physical address mapping information;
\circled{4} verify whether ``useful'' bit flips have occurred. If not, go to the step 2 or 3 to restart hammering based on the strategy;
\circled{5} gain the root/kernel privileges, say, by changing \emph{uid} to {\tt 0} for the current process.
The last two steps have been well studied~\cite{seaborn2015exploiting}. In the following, we describe the challenges of the first three steps.

\eat{
In the first step, we observe that there are many device-driver buffers that have the UaK ownership so as to speed up an I/O device's performance, and hence the buffers can be searched to identify the hammerable buffer. 
In the second step, we propose the memory-ambush technique to stealthily place the buffers alongside the page tables, as shown in Figure~\ref{fig:ambush}. This technique introduces low memory and page-cache overhead.
Note that we choose the page table as the victim object, as it is an appealing target~\cite{seaborn2015exploiting, van2016drammer}. 
Assuming that the objects have been successfully ambushed, we conduct an efficient single-sided rowhammer in the third step. This rowhammer algorithm builds from a timing-channel primitive~\cite{moscibroda2007memory}, which help filter out suitable address pairs. 
The last step is to verify whether expected bit flips have occurred. Specifically, if our exploit succeeds, then expected bit-flips has occurred within the page tables and the privilege has been escalated. Otherwise, another rowhammer attempt will be launched automatically. 
}

\mypara{Identify hammerable buffers} 
not all double-owned buffers are useful for our exploit.
A hammerable buffer should satisfy the following requirements: the buffer should be allocated from the kernel partition but can be accessed by unprivileged user processes. In addition, its size should be reasonably large (e.g., in the level of KB or MB). If it is too small, the number of bit flips could be considerably low. By imposing these constraints, our exploit is broadly applicable and potentially stealthy. 
    
\mypara{Stealthily position hammerable buffers and target objects} 
for rowhammer attacks to succeed, hammerable buffers and target objects must be physically adjacent to each other in the DRAM layout. Previous rowhammer attacks gaining the root/kernel privilege rely on technologies such as page deduplication~\cite{razavi2016flip} or exhausting either the page cache ~\cite{gruss2017another} or the system memory~\cite{gruss2016rowhammer, van2016drammer, tatar2018throwhammer} for this purpose. However, the page deduplication is usually disabled for security reasons~\cite{van2016drammer}, and other techniques are relatively easy to detect due to the anomaly in the page cache usage or the memory usage. As such, we need to design a new strategy that can position the hammerable buffers next to the target objects without exhausting the memory. 
    
\mypara{Efficiently perform hammering}
we cannot use double-sided hammering because the unprivileged user process no longer has access to the virtual-to-physical address mapping information or huge pages that are required to determine whether a pair of candidate addresses is separated by one row. On the other hand, the random hammering strategy of single-sided hammering could be inefficient. As such, we need to propose a new efficient hammer strategy without relying on the virtual-to-physical address mapping or huge pages.

\subsection{New Techniques} \label{sec:suboverview}
To address the aforementioned challenges, we present our main techniques as follows.

\subsubsection{Identification of Hammerable Buffers}\label{sec:hammerbuffer}
\begin{figure}[t]
\centering
\includegraphics[width=0.90\columnwidth]{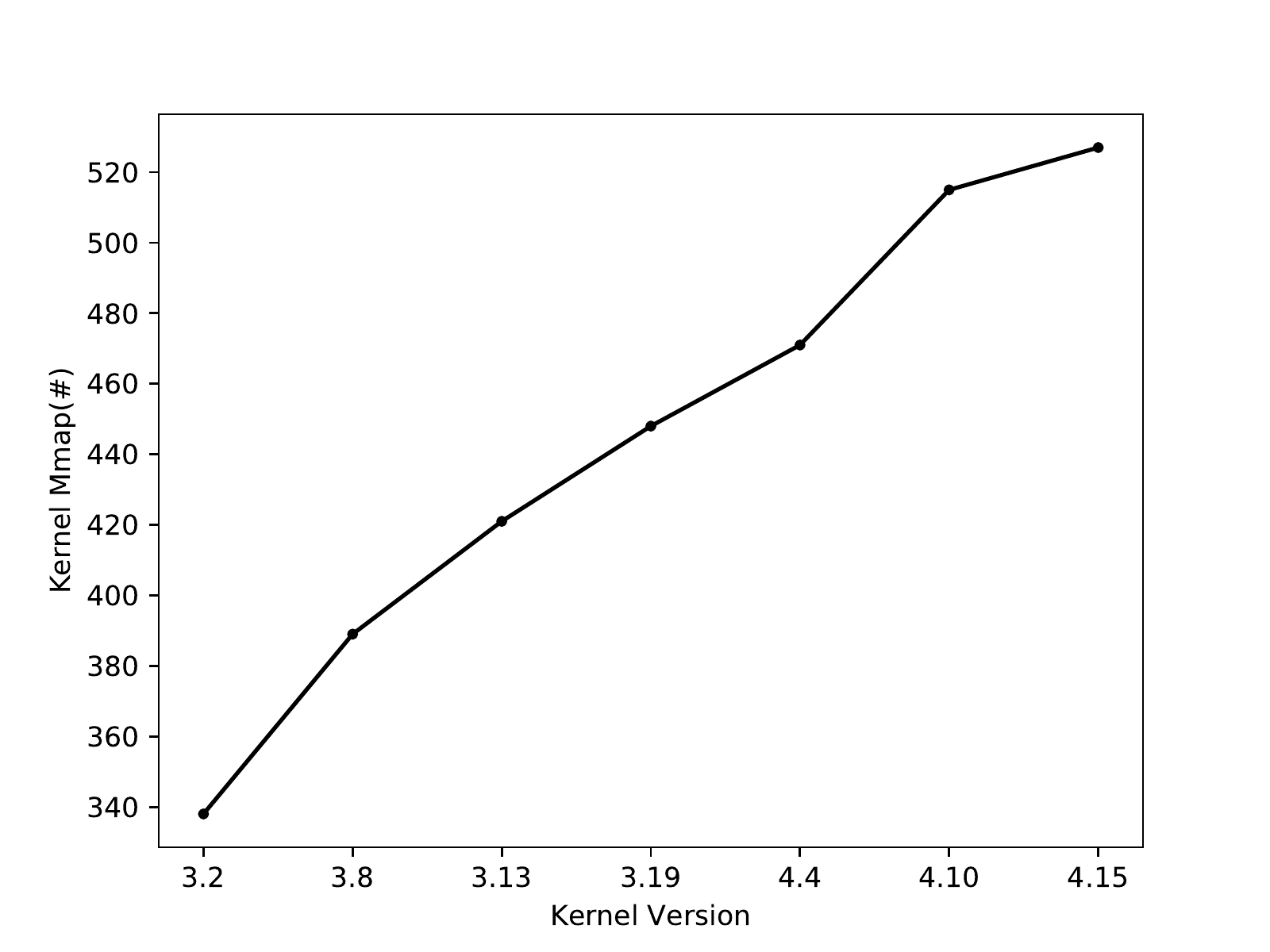}
\caption{The number of kernel \texttt{mmap} operations increases significantly as the Linux kernel evolves.}
\label{fig:mmap}
\end{figure}

\begin{figure}[t]
\centering
\includegraphics[width=0.90\columnwidth]{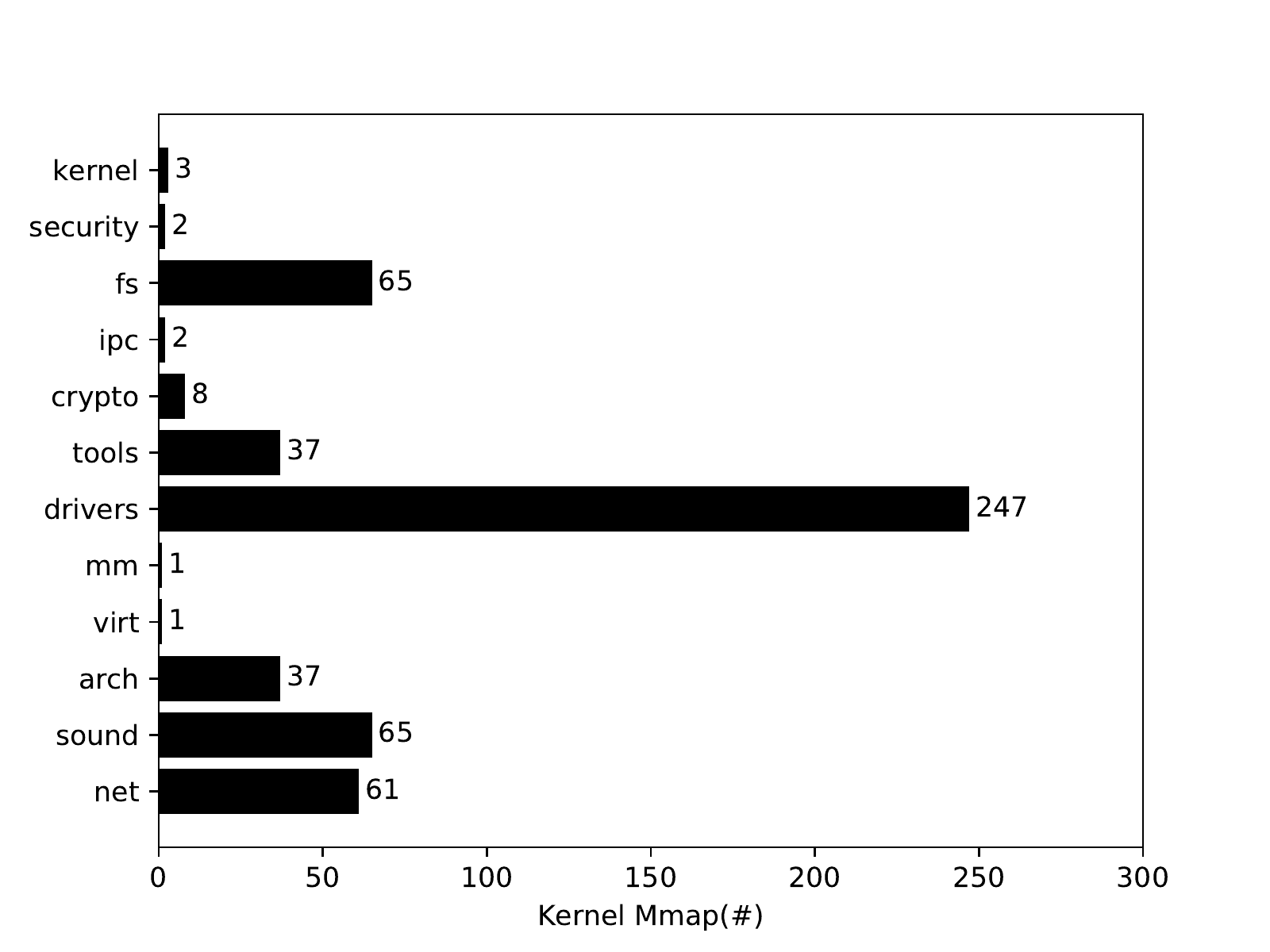}
\caption{The distribution of \texttt{mmap} operations in latest kernel version of $4.17.10$ (at the time of our experiments). Clearly, not only many device drivers in \texttt{drivers, sound, net} but also a small number of security-sensitive modules such as \texttt{security} and \texttt{crypto} also utilize the \texttt{mmap} feature for efficiency.}
\label{fig:dist}
\end{figure}

To eliminate the overhead of copying the kernel data into a user process and vice versa, the kernel supports mapping its own physical buffers into the user process through pre-defined interfaces (e.g., /dev). 
This efficient design requires the kernel to implement the memory-map {\texttt{mmap}} operation in the kernel space and provides the \texttt{mmap} system call in the user space. 
{From Fig.~\ref{fig:mmap}, we observe that the number of \texttt{mmap} operations inside the kernel increases rapidly as the Linux kernel grows (A kernel version shown in the figure is by default used in one release of Ubuntu Operating System). In a recent Linux kernel version of $4.17.10$, the total number grows up to $529$, indicating that the feature is applied to more and more kernel services.}
More specifically, the $529$ \texttt{mmap} operations are scattered over a dozen of kernel root directories such as \texttt{kernel, security, crypto} and \texttt{drivers}, shown in Fig.~\ref{fig:dist}. {Among the directories, \texttt{drivers, sound, net} take up the largest proportion, since numerous device drivers from devices such as SCSI, infiniband, graphics, Ethernet, media and Video4Linux have implemented the feature.}

Clearly, all such buffers are potential candidates for our exploit. Because the mmaped buffers are used by the kernel space, CATT accordingly allocates them from the kernel partition. This potentially allows a malicious user process to hammer the kernel after the buffers are mapped.
Arguably, CATT could revise its design and instead perform the allocation from the user partition. 
Intuitively, this modification exposes the kernel modules that utilize the mapped buffers to rowhammer.
Kernel modules implement \texttt{mmap} for different purposes. Some mmaped buffers are used in security-sensitive scenarios (e.g., \texttt{security} and \texttt{crypto}). If they are allocated from the user partition, they can be hammered by other user processes, introducing severe security threats. 
In addition, hardware devices most likely assume certain integrity of the data passed in from the drivers. 
As such, they would misbehave under rowhammer attacks.

As a result, both designs will make CATT inevitably susceptible to the rowhammer attacks. 
In this paper, we decide to evaluate the existing design made by CATT (i.e., allocate to-be-mapped buffers from the kernel partition). 
It would be an interesting future work to evaluate the security of the other option. 
Certainly, not all the mmapped buffers are exploitable to our exploit. 
We plan to design a program analysis system to help identify the hammerable buffers. 
{In our two proof-of-concept attacks, we have respectively selected the video buffers in the Video4Linux subsystem and the sg buffers in the SCSI subsystem for hammering.} These kernel buffers can be mmaped into the user process and thus become double-owned. They are allocated in a relatively large size (e.g., 18.75MB) and the buffers are virtually continuous but physically discontinuous, i.e., they can be mapped to any allocated physical pages.

\subsubsection{\upperambush} 
Our exploit uses double-owned video/sg buffers for hammering and targets the page tables. Therefore, we need to position video/sg buffers and page tables next to each other. 
To address that, we propose the \lowerambush technique to target page tables, which leverages the inherent design of the Linux kernel's {\tt mmap} and {\tt buddy} physical page allocator. We briefly introduce them first. 

\mypara{Mmap and page-table page allocation} {\tt mmap} is a posix API that allows a process to map files or devices into user-accessible memory. The caller of {\tt mmap} can specify the destination address, the source file descriptor, the protection, and a number of flags. 
For example, the {\tt MAP\_FIXED} flag requests the kernel to place the mapping at a specified address. This feature could be used to control the allocation of page-table pages. When a map is created, the kernel needs to populate the corresponding page tables and map the file/device (or the anonymous pages if the mapping is not backed by a file) at the selected addresses. However, this is usually done lazily, i.e., the page-table pages are not allocated or populated until the mapped addresses are accessed by the user process. Based on the above observations, we could make a \emph{mmap-based primitive} function, which takes a number as input and allocates page-table pages accordingly~\cite{seaborn2015exploiting}.

\begin{figure}[t]
\centering
\hspace*{-0.18in}
\includegraphics[scale=0.43]{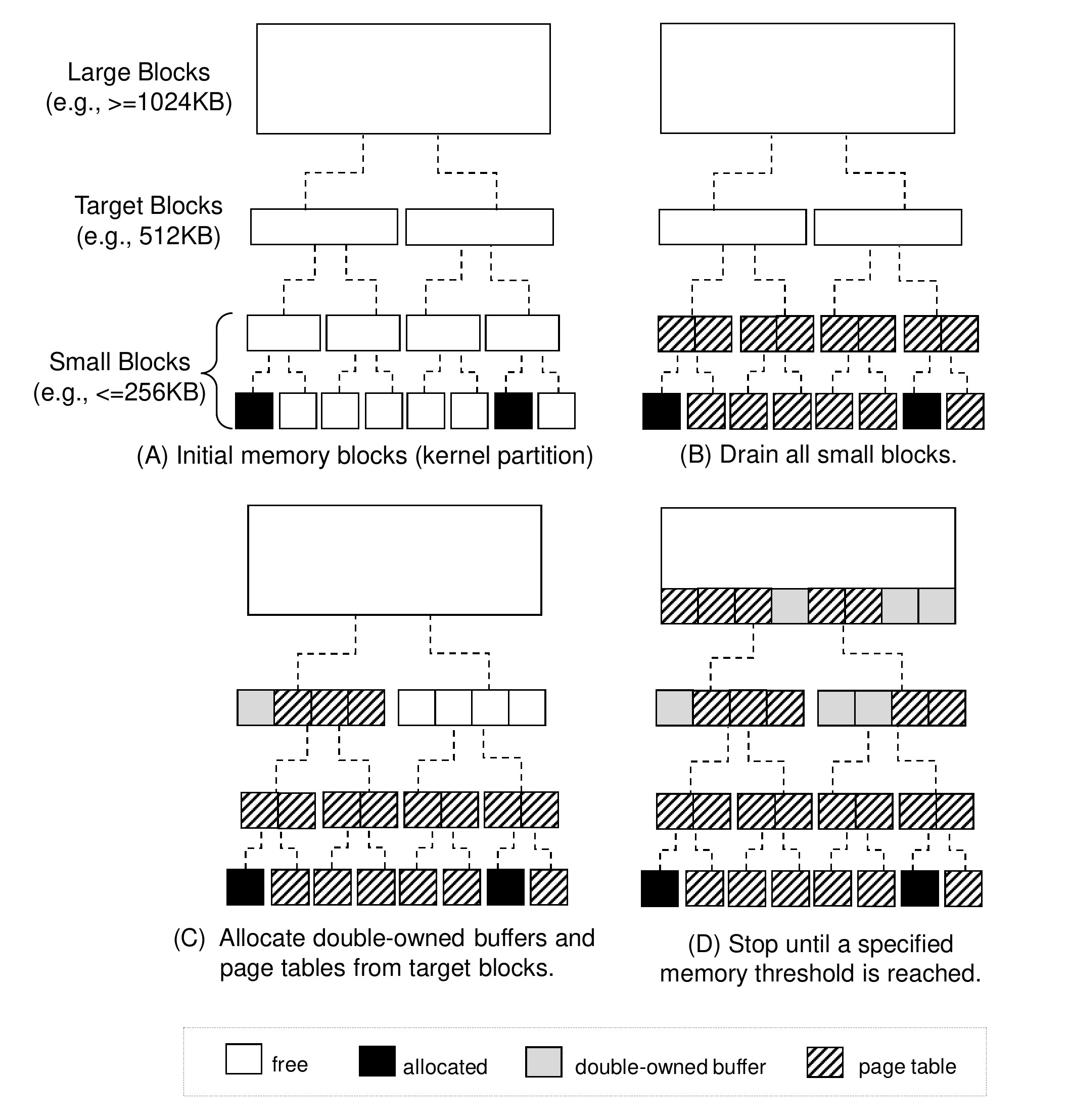} \\
\caption{\upperambush. {\textbf{(A)} shows the initial state of the Linux buddy allocator. The kernel memory is divided into blocks of different sizes and some blocks have been allocated (e.g., network ring buffer). In \textbf{(B)}, the {\tt mmap-based primitive} function is called repeatedly to fill the rest small blocks with page tables (Note that the memory-mapped \emph{tmp} file is allocated from the user partition). In \textbf{(C)}, a target block is split and allocated for double-owned buffers and page tables. \textbf{(D)} shows the state of the kernel partition after the previous step is repeated until a specified threshold is reached. (Note that the dashed line indicates that once a block is split into two equal blocks to satisfy a memory request, each split block has a smaller power-of-two number of pages and the unused split block will be linked to the block list that has the same number of pages.)}}
\label{fig:ambush}
\end{figure}

\mypara{Linux buddy allocator}
like most OS kernels, the Linux kernel uses layers of memory allocators to fulfill different needs of the kernel. In particular, the physical pages are allocated using the buddy allocator~\cite{gorman2004understanding}. As shown in Fig.~\ref{fig:ambush}\textbf{(A)}, the buddy allocator splits memory into equal halves called blocks. Each block initially contains a power-of-two number of pages that are physically continuous. 
Upon an allocation request, the kernel searches the blocks that best match the request. If the blocks do not have enough continuous pages for the request, the kernel splits a larger block in half and returns one half to the request. This process can happen recursively. For example, to allocate $256$KB of memory, the buddy allocator will first search the blocks that contain 64 pages (a single page is $4$KB). If none is found, the kernel tries to split a large block of $512$KB ($128$ pages) in half to fulfill the request. 
When the requested pages are freed, the kernel tries to merge them with other free pages if possible. To continue the previous example, if the allocated 64-page memory is freed, the kernel checks whether its buddy (i.e., the other $64$-page memory in the $512$KB block) is free. If so, the kernel merges them to recreate the split large block.

\mypara{\upperambush}
{as mentioned above, the main purpose of \lowerambush is to position the double-owned buffers next to the page tables by leveraging the Linux features. It is relatively easy to create multiple page-table pages in Linux, for example, by repetitively invoking \emph{mmap-based primitive} function to map the same user file into different parts of the user address space. When a page-table page is created, the kernel will ask the buddy allocator to allocate a 4KB-block (x86-64 also supports large page sizes, such as 2MB and 1GB).}
 
{Note that, the two objects (i.e., the buffers and the page tables) that are physically consecutive may not be adjacent to each other in the memory, because the mapping between the physical address and the DRAM layout is not linear~\cite{brasser17can}. Some bits of a physical address are used to select the DIMM, Rank, Bank and row. 
For example, our test machine has two DIMMs and the 6th bit selects between the two DIMMs. The consecutive physical
addresses such as 0x1000000 and 0x0FFFFFF are located on different DIMMs and thus are not next to each other.}

{To address this challenge, we need to find certain blocks that can occupy two adjacent rows in the DRAM layout (one row for double-owned buffers and the other one for page tables).
Since the mapping of physical addresses to DRAM channels, ranks, banks and rows has been reverse engineered~\cite{xiao2016one,pessl2016drama}, the row index that a physical address maps to is determined by the most significant bits of the physical address. 
For instance, physical address bits from b$_1$$_8$ to b$_3$$_2$ on Sandy Bridge, b$_1$$_8$ to b$_3$$_1$ on Ivy Bridge, b$_2$$_3$ to b$_3$$_4$ on Haswell, respectively decide the DRAM row index. Thus, a block of row size per row index occupies one row entirely since the block's starting physical address is row-aligned (e.g., 0x40000).}
To this end, the size of a target block (i.e., \emph{TargetBlockSize}) should be \emph{twice} of the row size (shown in Equation~\ref{equ:targetbs}).
The row size (i.e., \emph{RowsSizePerRowIndex})  is determined by the number of DIMMs, the number of banks and the size of a single row in one bank (shown in Equations~\ref{equ:rowsize} and \ref{equ:rows}).
The specific equations are listed as follows:
\begin{small}
\begin{equation}
\mathit{TargetBlockSize = RowsSizePerRowIndex \cdot 2} \label{equ:targetbs} 
\end{equation}
\end{small}
\begin{small}
\begin{equation}
\mathit{RowsSizePerRowIndex = DIMMs\cdot BanksPerDIMM\cdot RowSize} \label{equ:rowsize}
\end{equation}
\end{small}
\begin{small}
\begin{equation}
\mathit{BanksPerDIMM = BanksPerRank \cdot RanksPerDIMM} \label{equ:rows}
\end{equation}
\end{small}

The memory ambush technique is illustrated in Fig.~\ref{fig:ambush}. 
Specifically, the blocks smaller than the target blocks are \emph{small blocks}, and the blocks larger than the target blocks are \emph{large blocks}.
On our test machine, the row size (i.e., \emph{RowsSizePerRowIndex}) is 256KB, and the size of the target block (i.e., \emph{TargetBlockSize}) is 512KB. 

At the beginning of our technique, the memory of the kernel partition could be fragmented, especially for small blocks (Fig.~\ref{fig:ambush}\textbf{(A))}. Next, we drain the small blocks by repeatedly invoking \emph{mmap-based primitive} function to allocate page-table pages (Fig.~\ref{fig:ambush}\textbf{(B)}). We check the depletion of small blocks by accessing the file of \texttt{/proc/buddyinfo}. Note that any process, privileged or not, can read this file to obtain data about the available and allocated blocks. 

Based on the obtained data, we can then position the double-owned buffers next to the page tables, since they are expected to share the same target block (Fig.~\ref{fig:ambush}\textbf{(C)}).
Note that the double-owned buffers usually have a limited/fixed size. If the size happens to be one or more of \emph{RowsSizePerRowIndex}, the two objects share the target block equally. 
If the buffers cannot stuff one row or there is a remainder after stuffing one or multiple rows, the page tables ought to occupy the remaining empty pages of the split target block.
We repeat this step until a specified memory threshold is reached (Fig.~\ref{fig:ambush}\textbf{(D)}). By doing so, we can stuff the empty pages of the split target block, and position more page-table pages next to the buffer pages, increasing the probability that the buffers are in the aggressor rows while the page tables stuff the victim rows. 

{In our experiments, we have two machines. For one machine, we keep the initial Linux system running a typical workload (i.e., a browser, a mail client, and a music player). As such, the depleted small blocks are calculated to be $56$MB. The other machine launches not only the above typical workload, but also the office suite. The corresponding small block size is $115$MB.
Both workloads take up a small part of the whole system memory. In addition, stuffing the small blocks with page tables might also increase the chance of positioning the two objects next to each other.}
Certainly, the access to \texttt{/proc/buddyinfo} can be removed or protected without causing problems for most programs. We argue that this is similar to previous systems that use {\tt pagemap} to obtain virtual-to-physical address mapping: the access to this file is currently enabled by default and thus can be misused by any one. 

\mypara{Distinguishing \lowerambush} 
{Clearly, our technique is quite different from memory spray~\cite{seaborn2015exploiting} and memory waylay~\cite{gruss2017another}. We list the major differences between \lowerambush and memory groom ~\cite{van2016drammer}.
Firstly, the memory groom technique (which is termed as \texttt{Phys Feng Shui})
deterministically places a page-table page into an attacker-chosen, vulnerable DRAM row. To this end, \texttt{Phys Feng Shui} firstly requests all large physically-continuous blocks (each block size is greater than the DRAM row size) and performs a memory scan to target a vulnerable row. After that, it requests all other medium blocks (equal to the row size). At this moment, it is highly likely to cause an out-of-memory (OOM) situation. In our experiments, the required memory using \texttt{Phys Feng Shui} is $7.62$GB, taking up 99.3\% of the total available memory (i.e., $7.7$GB) and thus triggers a system crash. This observation is also confirmed in Figure 2 of~\cite{gruss2017another}.
To relax the memory requirement and maintain the system stability, we only repeatedly request $4$KB-sized memory blocks using the \emph{mmap-based primitive} function, until a predefined memory threshold is reached. The threshold is safe enough to avoid the OOM situation.}

{On top of that, we explicitly request blocks of $4$KB, which implicitly forces the allocator to split blocks larger than $4$KB. This consumption strategy is ``small blocks first".
For \texttt{Phys Feng Shui}, it explicitly forces the allocator to allocate memory from large, medium to page-sized blocks, which we term ``large blocks first".}

{Further, the essential reason behind the OOM situation is that the Linux buddy allocator by default avoids placing kernel objects (page tables in this scenario) near userspace objects and it only deviates from this default behavior in a near-OOM situation. As a result, both \texttt{Phys Feng Shui} and the memory spray technique~\cite{seaborn2015exploiting} require memory exhaustion so as to place the page-table page of kernel space next to attacker-controlled pages of user space.
In contrast, we conform to the default behavior of the allocator. The attacker-controlled pages that we request also reside in the kernel space; thus we are able to ambush the page-tables pages with only a small amount of memory.}

{Lastly, \texttt{Phys Feng Shui} does not leverage the DRAM mapping function, which is important to our technique. 
As mentioned above, a block of row size per row index occupies one row entirely. As such, we always force the allocator to drain available small blocks and then split the target block to satisfy the double-owned buffer allocation and then the page-table allocation. By doing so, the two objects can be adjacent to each other in the DRAM memory.}

\subsubsection{Efficiently Hammering}~\label{sec:efficienthammer}
Since Linux kernel 4.0, the access to {\tt pagemap} has been protected from unprivileged processes. Without the information about the virtual-to-physical address mapping, an intuitive solution is to randomly select a pair of virtual addresses to hammer, also known as the single-sided rowhammer. Overall, this approach is less effective than double-sided hammering (e.g., if these two addresses happen to lie in the same row). In our system, we resort to a timing channel~\cite{moscibroda2007memory} to improve the efficiency of single-sided hammering.  

{Specifically, this timing channel is created by the row-buffer conflicts within the same DRAM bank. As we previously mentioned, each bank has a row buffer that caches the last accessed row. If a pair of virtual addresses reside in two different rows of the bank and they are accessed alternately, the row buffer will be repeatedly reloaded and cleared. This causes the so-called row-buffer conflicts. Clearly, row buffer conflicts can lead to higher latency in accessing the two addresses than the case that they lie either within the same row or in different banks. As such, we are highly likely to distinguish whether two addresses are in different rows within the same bank. When we perform the hammering, we can select such pairs of addresses as candidates, thus improving the efficiency of the single-sided hammering. Note that there are previous works~\cite{bhattacharya2016curious,xiao2016one,pessl2016drama,van2018guardion,seaborn2015exploiting} that also exploit this timing channel for different purposes. For instance, Bhattacharya et al.~\cite{bhattacharya2016curious} leveraged this channel to determine which DRAM bank that the target secret exponent resides in.}

\section{Proof-of-concept Attacks}\label{sec:impl}
In this section, we present in detail two proof-of-concept attacks that exploit two different double-owned buffers to break the kernel-user separation enforced by CATT. {At a high level, our attacks respectively use the double-owned video buffers and SCSI generic buffers as the aggressor rows and targets the page table.} Both attacks then rely on our \lowerambush technique to stealthily position the aggressor and victim rows adjacent to each other. Furthermore, the attacks perform the improved single-sided hammering. We also briefly describe the steps to verify whether the attacks have succeeded and to gain the root and kernel privileges if so.  

\subsection{Double-owned Video Buffers}
Video4Linux (V4L) is a collection of device drivers that provide the API for programs to capture real-time videos on the Linux systems. The current specification of V4L is version2 (V4L2). A V4L device is usually presented as a char device under the directory {\tt /dev} of the file system, such as {\tt /dev/video0}. The device is accessible by unprivileged processes by default. 

We dive into the memory allocation by the V4L2 device and discover that the video buffer is allocated by the kernel and mapped into the user space.
Specifically, by issuing the \texttt{VIDIOC\_REQBUFS} {\tt ioctl} command, an unprivileged process can request the V4L2 driver to allocate physical device memory as the video buffers.  
Note that CATT will allocate this block of device memory from the kernel partition~\footnote{Alternatively, CATT can be extended to allocate this block of memory from the user partition. For the potential security reasons stated in Section~\ref{sec:hammerbuffer}, we follow the design of the current CATT system.}. 
When the request is completed, the unprivileged process can then calls the \texttt{mmap} function to map the allocated memory into its own address space with read and write permissions. Accordingly, the \texttt{uvc\_v4l2\_mmap} function inside the device driver will be invoked to actually perform the mapping. 
Until now, the video buffers are changed to be double-owned buffers, facilitating the unprivileged user process to hammer the kernel.
The maximum size of the video buffer for a V4L device is limited to $18.75$MB, a sufficient size for our attack. 

In the following, we briefly summarize the five steps for an unprivileged process to obtain read access to the video buffers: 

\begin{itemize}[itemsep=1ex,leftmargin=0.4cm]
    \item \emph{Open the video device:} the V4L2 video capture device is a char device (as opposite to a block device) located in the \texttt{/dev} directory. Linux can support up to 64 V4L2 devices, starting from {\tt /dev/video0} to {\tt /dev/video63} with a major number of $81$ and a minor number from $0$ to $63$. We select  {\tt /dev/video0}  as our device. 
    \item \emph{Configure the video device:} different video capture devices support different capabilities, such as cropping limits, the pixel aspect of images, and the stream data format. We apply the default settings to this device.
    \item \emph{Request the video buffer:} after the configuration, we can issue the \texttt{VIDIOC\_REQBUFS} command to ask the driver to allocate the video buffer.  The command provides three ways for a user process to access the allocated kernel memory, i.e., memory mapped, user pointer, or DMABUF based I/O~\cite{v4l2}. Moreover, it allows the process to request up to $32$ buffers and the size of each buffer is $600$KB (i.e., $18.75$MB in total). For our attack, we specify the memory mapped I/O and use the maximum buffer size. 
    \item \emph{Map the video buffer:} the \texttt{VIDIOC\_QUERYBUF} command returns the detailed information about the allocated video buffers (e.g., the size and address of each buffer set). Based on this information, we can map all these buffers into the user space. 
    \item \emph{Close the video device:} after our rowhammer exploit completes, we should unmap the video buffers from the user space and close the video device. 
\end{itemize}

\subsection{Double-owned SCSI Generic (sg) Buffers}
{The SCSI Generic packet device driver (sg) is one of the four high level SCSI device drivers along with sd (direct-access devices-disks), st (tapes) and sr (data CDROM). The sg is a char device while the other three are block devices.  
The sg driver is able to find $256$ SCSI devices and it is often used for scanners, cd writers and reading audio cds. 
An sg device is located under the directory {\tt /dev} such as {\tt /dev/sg1}. An unprivileged user is also capable of accessing the device with read and write permissions.}

{Similar to the video buffer, the sg buffer is a kernel buffer. Upon each {\tt open} request, CATT will allocate an sg buffer from the kernel partition. To remove the extra data copy between kernel and user spaces, the sg driver also provides the memory-mapped IO interface to map the allocated sg buffer into the user space. As such, an unprivileged process has four steps to make the sg buffer double-owned.}

{\begin{itemize}[itemsep=1ex,leftmargin=0.4cm]
    \item \emph{Open the sg device:} as sg is a char-based Linux device driver, it provides basic {\tt open/close/ioctl} type interfaces. Unlike the video buffer, the sg buffer will be automatically reserved when the device is opened (we choose {\tt /dev/sg1} to open). 
    \item \emph{Request the sg buffer:} the default size of the reserved sg buffer is only $32$KB and cannot even stuff one row, an insufficient size for our attack. To address this issue, we can issue the \texttt{SG\_SET\_RESERVED\_SIZE} command using {\tt ioctl} to increase its size up to $124$KB. Clearly, such sg buffer size is still not enough. We observe that the device can open for multiple times and a dedicated sg buffer will be allocated for each open. As the default open-file limit for a non-root user is $1024$, then the maximum number of open files for the particular sg device is $1021$ (with {\tt stdin, stdout, stderr} excluded). As such, the maximum sg buffer size can be increased to $123.64$MB, much larger than that of the video buffer. For our attack, only $31$MB (i.e., $256$ * $124$KB) is enough. 
    \item \emph{Map the sg buffer:} based on the device file descriptor returned by the device open and the sg buffer size returned by the \texttt{SG\_SET\_RESERVED\_SIZE} command, we can successfully map the sg buffer into the user space for every open device file descriptor. On top of that, we need to issue the \texttt{SG\_FLAG\_MMAP\_IO} command through {\tt ioctl} before we proceed to procure the allocated sg buffer.
    \item \emph{Close the sg device:} after our rowhammer exploit completes, we gracefully unmap the sg buffers from the user space and close the device. 
\end{itemize}
}

\begin{algorithm}[t]
	\caption{\upperambush}\label{alg:ambush}
	\begin{algorithmic}[1]
		\If{$video$ $ $ $is$ $ $ $defined$}
		\State $dev\_buf\_size \gets \textit{18.75MB}$
		\Else
		\State $dev\_buf\_size \gets \textit{31MB}$
		\EndIf
		\State $file\_size \gets \textit{2MB}$
		\State $page\_size \gets \textit{4KB}$
		
		\Loop
		\State $pt\_size \gets \textit{(threshold\_mem\_size - dev\_buf\_size) - file\_size}$
		\State $map\_mem\_size \gets \textit{pt\_size * 512}$     
		\State $vma\_num \gets \textit{map\_mem\_size / file\_size}$  
		\If {$vma\_num < \textit{VMA\_limit}$} 
		\State $break$     
		\EndIf
		\State $file\_size \gets \textit{file\_size * 2}$     
		\State \textbf{goto} \emph{loop}
		\EndLoop
		\State $map\_mem\_base \gets \textit{mmap(map\_mem\_size)}$   
		\State $file \gets \textit{create(file\_size, file\_path)}$   
		\State $map\_each\_base \gets \textit{map\_mem\_base}$
		\State $pt\_size\_each\_while \gets \textit{(file\_size / page\_size * 8)}$ 
		\State $pt\_size\_sum \gets \textit{0}$
		\State $idx \gets \textit{0}$
		\State $small\_blocks\_size \gets \textit{/proc/buddyinfo}$
		\While{$idx < vma\_num$}
		\State $mmap(map\_each\_base, file\_size, file)$
		\State $read\_access(map\_each\_base, file\_size)$
		\State $map\_each\_base \gets \textit{map\_each\_base + file\_size}$
		\State $pt\_size\_sum \gets \textit{pt\_size\_each\_while + pt\_size\_sum}$
		\If {$idx == 0$}  
		\State \textit{add\_marker\_to\_each\_file\_page\_header()}
		\EndIf 
		\If {$pt\_size\_sum == small\_blocks\_size$}  
		\State \textit{dev\_buf\_allocation(dev\_buf\_size)}
		\EndIf 
		\State $idx \gets \textit{idx + 1}$
		\EndWhile
	\end{algorithmic}
\end{algorithm}

\subsection{\upperambush}
In this technique, we need to create sufficient page-table pages under the given targeted memory. 
Specifically, we create a temporary file \emph{tmp} using \texttt{tmpfs}, which is stored in the memory only. We then map this file repeatedly in order to create numerous virtual memory areas (VMAs) mapped to the file. In each call to \texttt{mmap}, all pages in the mapped area are accessed in order to populate the page table.  
The size of this file needs some careful considerations: it should not be too large to avoid an excessive usage of the physical memory; otherwise we risk being detected; the size should not be too small because Linux limits the number of VMAs that can be created by {\tt mmap} (i.e., $65536$). We need a sufficiently large number of page-table pages to be targeted for the attacks to succeed. 
 
The size of the \emph{tmp} file is calculated in line 6 to line 17 of Algorithm~\ref{alg:ambush}. Specifically, the file size is initialized to $2$MB in line 6, which can be mapped by a single PTE (page table entry) page~\footnote{Each PTE page is 4KB and each PTE is 8B. Therefore, one PTE page consists of 512 entries and each entry maps a 4KB virtual address to the physical address, which totals to 2MB.}. In line 9 to 11, we calculate how many VMAs we need to create if we were to use a specified memory threshold for the attacks. More specifically, line 9 calculates the size of the page tables that can be created. Note that {\tt dev\_buf\_size} is the size of double-owned buffer that we can request and {\tt threshold\_mem\_size} is the total memory specified for the attacks. It is a user configurable parameter.
In our experiment,  the {\tt threshold\_mem\_size} can be as low as 88MB.
Line 10 calculates how much memory can be mapped by these page-table pages; Line 11 calculates the number of VMAs to be created. If the number is less than the limit, we have found the right file size. Otherwise, we double the file size and try again. 

Based on \texttt{vma\_num}, the \emph{tmp} file is mapped and accessed repeatedly to indirectly populate many created PTE pages (line 25-37). In the first iteration, we place a special marker in every page of the \emph{tmp} file. Since the file is mapped in all the locations, we can look for this mark to check whether the bit flips caused by rowhammer have changed the page table, i.e., whether the attacks succeeds or not. 
When all the small blocks are drained, we start to call the {\tt dev\_buf\_allocation} function. 

{For the video buffer, the function returns $32$ buffers (size of each buffer is $600$KB). As such, $32$ large blocks of $1024$KB will be allocated and shared by the video buffers and subsequent PTE pages. 
For the sg buffer, $256$ buffers (each buffer is $124$KB) will be returned, which will share $256$ memory blocks (target blocks first and then a few large blocks if the target blocks are depleted) with the PTE pages. 
Given the size of all rows per row index is 256KB, each video buffer crosses three consecutive row indices, stuffing the first two rows and leaving the last row partially occupied. The last row is then stuffed by the PTE pages, neighboring one row of the video buffers.  
For the sg buffer, each buffer will occupy one row partially while the PTE pages stuff the row's remaining part and one adjacent row.}

\subsection{Efficient Single-sided Hammering}
Since we do not have access to the virtual-to-physical address mapping information, we rely on the single-sided hammering but improve its efficiency with the timing channel based on the row buffer. As mentioned in Section~\ref{sec:efficienthammer}, a pair of virtual addresses in different rows of the same bank has longer access latency than these in the same row or in the different banks. Such pair will be selected to hammer. For brevity, we call such pair of addresses DRSB (different rows within same bank). 

{On both machines, most pairs of virtual addresses in DRSB have an access latency that is no less than $360$ elapsed CPU cycles (see Section~\ref{sec:eva}). 
When a pair of virtual addresses from the video/sg buffers is randomly selected, we will time its latency. If it is no less than $360$ CPU cycles, then we use the pair for hammering. Otherwise, the pair will be discarded. We repeat the step until the attacks succeeds.}

\begin{algorithm}[t]
	\caption{Verification and Privilege Escalation}\label{alg:search} 
	\begin{algorithmic}[1]
		\State $page\_size \gets \textit{4KB}$
		\State $idx1 \gets \textit{0}$
		\While {$idx1 < map\_mem\_size / page\_size$} 
		\LineComment{map\_mem\_base is the mapped base address.} 
		\State $ptr1 \gets \textit{map\_mem\_base + idx1 * page\_size}$ 
		\State $val1 \gets \textit{(map\_mem\_base + idx1 * page\_size)}$ 
		\LineComment{check whether the page is a controllable page table.}
		\If {$val1 \not= \textit{marker}$} 
		\LineComment{save the second entry of the page table.}
		\State  $old\_pte \gets \textit{ptr1[1]}$ 
		\LineComment{set the physical page \#0 readable and writable.}
		\State  $new\_pte \gets \textit{0x27}$ 
		\State  $idx2 \gets \textit{1}$ 
		\While{$idx2 < map\_mem\_size / page\_size$}
		 \LineComment{pick the second virtual page.}
		\State $ptr2 \gets \textit{map\_mem\_base + idx2 * page\_size}$
		\State $val2 \gets \textit{(map\_mem\_base + idx2 * page\_size)}$ 
		\LineComment{find out the target page table.}
		\If {$val2 \not= \textit{marker}$ \textbf{and} $idx2 \not= idx1$} 
		\State \textit{privilege\_escalation(ptr1, ptr2)}
		\State $return$
		\EndIf 
		\LineComment{each page-table page has 512 entries.}
		\State $idx2 \gets \textit{idx2 + 512}$ 
		\EndWhile
		\State $ptr1[1] \gets \textit{old\_pte}$
		\EndIf
		\State $idx1 \gets \textit{idx1 + 1}$
		\EndWhile
	\end{algorithmic}
\end{algorithm}

\begin{figure*}[ht]
	\centering
	\begin{subfigure}[t]{0.48\textwidth}
		\centering
		\includegraphics[height=2.55in]{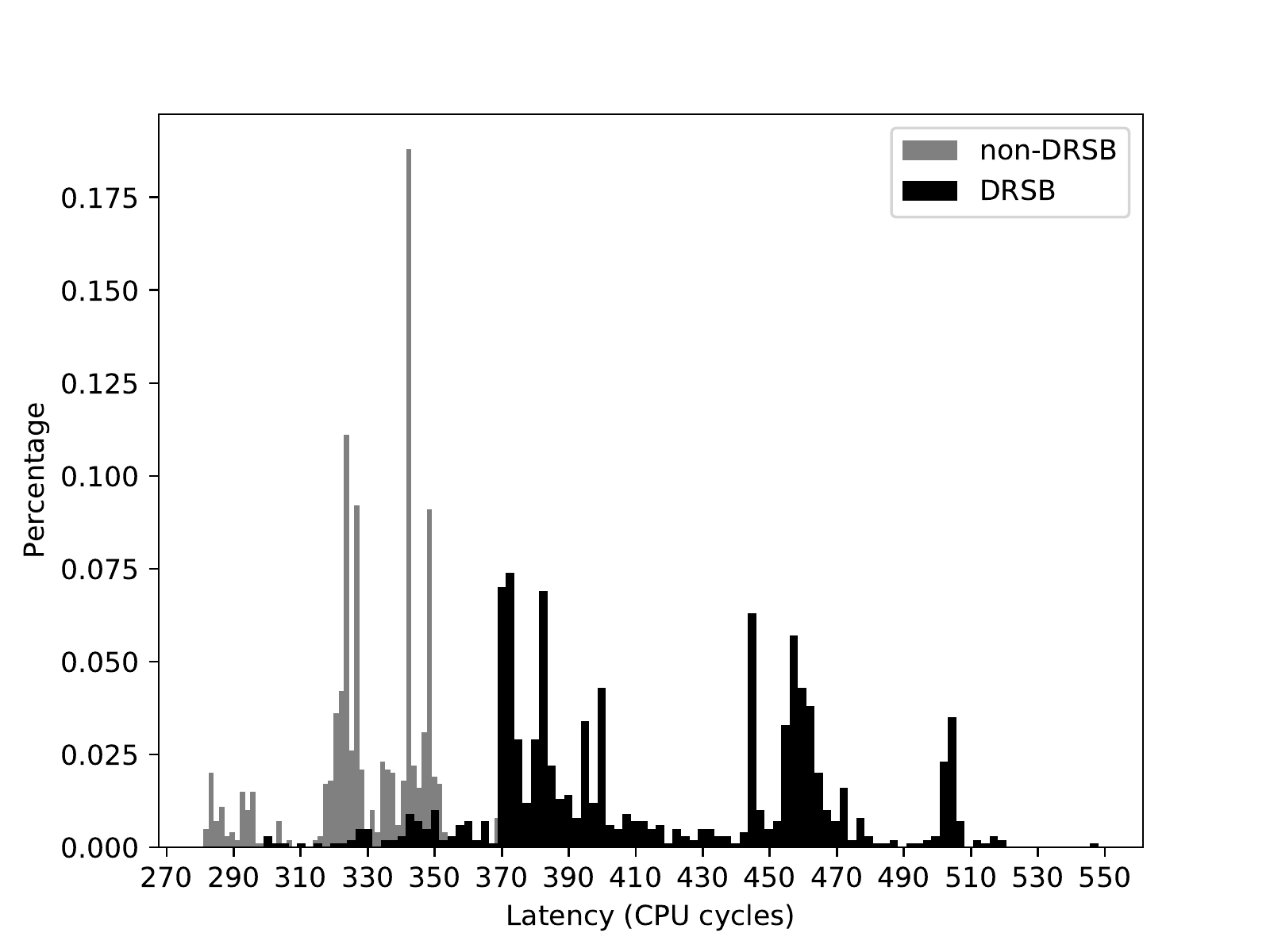}
		\caption{{For the Dell, $927$ out of $1000$ pairs in DRSB have access latency of no less than $360$ CPU cycles. In contrast, $974$ out of $1000$ pairs in non-DRSB have access latency  of less than $360$ CPU cycles.}}
		\label{fig:dell}
	\end{subfigure}
	~
	\begin{subfigure}[t]{0.48\textwidth}
		\centering
		\includegraphics[height=2.55in]{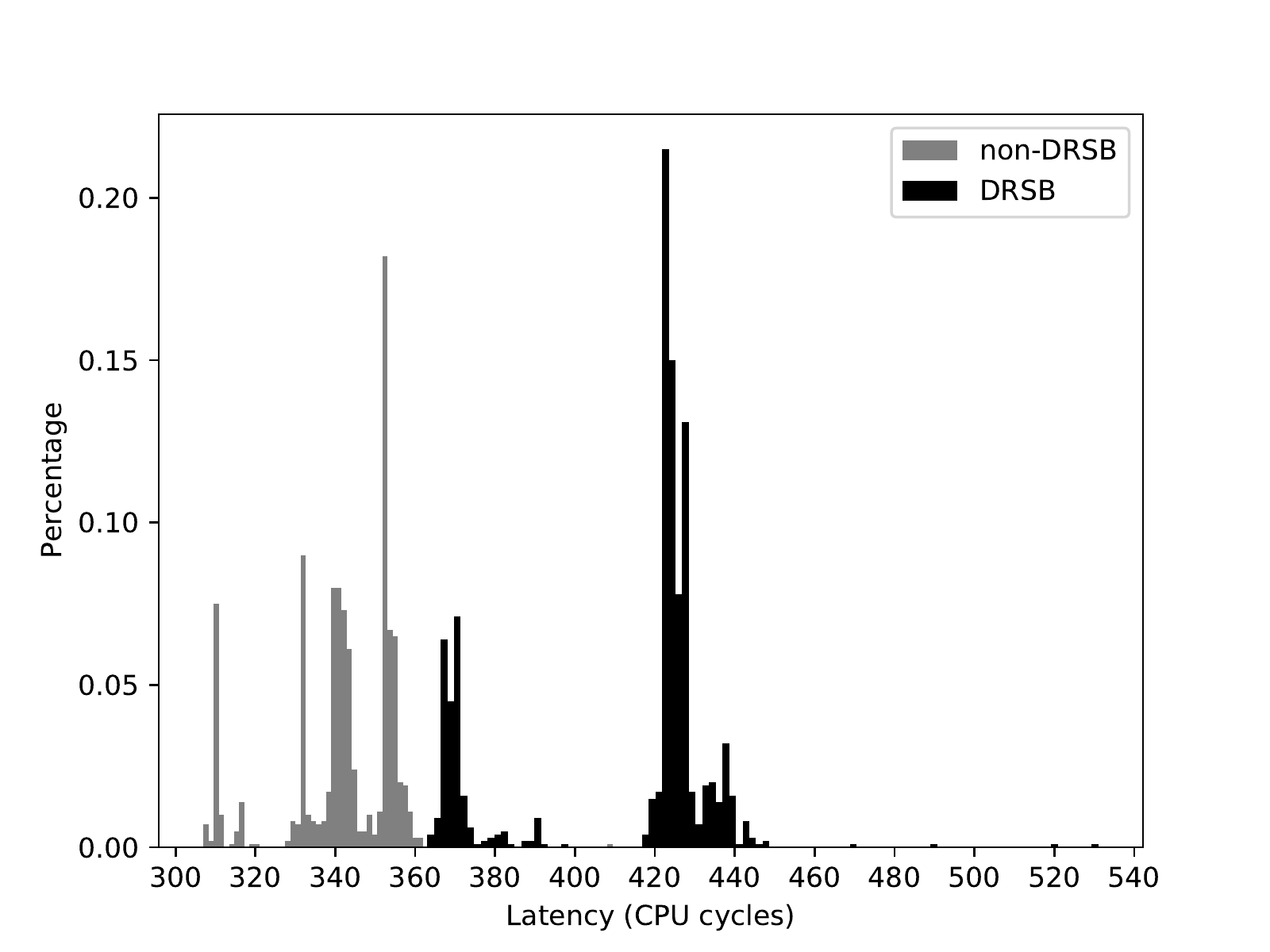}
		\caption{{For the Lenovo, $1000$ pairs in DRSB have access latency of more than $360$ CPU cycles. In contrast, $990$ out of $1000$ pairs in non-DRSB have access latency of less than $360$ CPU cycles.}}
		\label{fig:lenovo}
	\end{subfigure}%
	\caption{
	{For each machine, most pairs in different rows within same bank (DRSB) have longer access latency than that of most pairs in non-DRSB.}}
	\label{fig:wosbdr}
\end{figure*}

\subsection{Privilege Escalation}
In our attacks, we randomly select a pair of addresses in DRSB for hammering. We then check whether the attacks succeeds and select a new pair if not. Our attacks target the page tables. We aim at hammering the video/sg buffers to flip bits in adjacent page tables. If the bit flips happen to change the mapped physical address to a page table, we can gain full control over all the system memory. This is feasible because we (lightly) spray the kernel memory with page tables. This process is described in detail in Algorithm~\ref{alg:search}. 


{As previously mentioned, we embed a special marker at the beginning of every mapped page. We can check for this marker to tell whether the hammering has caused the address to be remapped. Specifically,  after each round of hammering, we read every page-aligned virtual address mapped to the \emph{tmp} file and check whether the returned value (i.e., \texttt{val1}) is equal to the marker  (\emph{line 6}). If they are equal, we continue to check the next page for success.
Otherwise, we have found a virtual page $V_a$ that points to a physical page outside of the \emph{tmp} file due to the bit flips. Next, we need to check if the page $V_a$ itself is a writable page-table page (\emph{line 10-18}). To this end, we  pretend that $V_a$  is a writable page table and tentatively modify one of the entry. We then read all the markers again. If another virtual page $V_b$ has been remapped, page $V_a$ is an attacker-controlled page table, and we can use $V_b$ to access the maliciously mapped memory. 
Now that we have read-write access to any system memory,  we essentially have gained full control over the system (i.e., \emph{kernel privilege}).}

{We also try to change the \texttt{uid} of current user process to \texttt{0} to gain the root privilege. Without the access to \texttt{pagemap}, we have to scan all the available physical memory to locate current process's credential structure, struct {\tt cred} that stores the critical \texttt{uid} field. 
To make this search fast and precise, we construct a distinct pattern in the {\tt cred} structure and search each page for this pattern. Specifically, the {\tt cred} structure contains three user ids (e.g., \texttt{uid} and \texttt{suid}) and three group ids (e.g., \texttt{gid} and \texttt{sgid}) stored sequentially. We firstly set the three user ids and group ids to be the same as \texttt{uid} using syscalls such as {\tt seteuid}. We can use these ids as a pattern to search for our {\tt cred} structure. Specifically, in a loop, we set a page table entry in page $V_a$ to every available physical page and scan the newly mapped page to check whether it contains the pattern. Once the pattern is located, we overwrite \texttt{uid} to \texttt{0} and then invoke {\tt getuid} function to verify whether the \texttt{uid} of current process has been changed to 0. If not, there must be another user process that has the same uid fields, so the revised {\tt cred} structure is restored and the verification continues. This essentially gives the root privilege. Note that after each change to the address translation in $V_a$ we need to ensure that the CPU's TLB is reloaded; otherwise the CPU will continue using the old address translation. However, a user process does not have the privilege to flush the TLB. To address this problem, we flush the CPU cache with the \texttt{clflush} instruction to ensure that the change to the page table is committed to the memory and then schedule the attacking process between two physical cores. 
In the meantime, two other user processes have been running on the two cores, respectively and they do nothing in an infinite loop. 
When the attacking process is scheduled onto either core, context switch will occur and thus the TLB is flushed automatically.}

\subsection{Evaluation}\label{sec:eva}
In this section, we firstly describe how to measure the memory access latency required for the efficient hammering, then evaluate the effectiveness and stealthiness of our attacks. All the experiments were conducted on two machines (i.e., 
Dell Latitude E6420 PC with 2.8GHz Intel Core i7-2640M and 8GB DDR3 memory and Lenovo Thinkpad T420 PC with 2.6GHz Intel Core i5 2540M and 8GB DDR3 memory. The operating system running above each machine is Ubuntu 16.04 LTS for x86-64. Dell E6420 has a Linux kernel of 4.10.0-generic while Lenovo T420 has a Linux kernel of 4.8.0-generic. Note that both machines are vulnerable to the rowhammer bug.

\begin{table*}
\renewcommand\arraystretch{2.0}
\centering
\begin{tabular}{cccccccccc}
\hline
\multirow{2}{*}{\textbf{Machine Type}} & \multirow{2}{*}{\textbf{Occupied Memory}} & \multicolumn{2}{c}{\textbf{Exploitable Buffer}} & \multicolumn{2}{c}{\textbf{{\upperambush}}} & \multicolumn{2}{c}{\textbf{Flippable Bit Occurs}} & \multicolumn{2}{c}{\textbf{Exploitable Bit Occurs}} \\
& & \quad {Type} & Size & \# of Runs & S.R. & \# of Runs & S.R. & \# of Runs & S.R. \\ 
\hline
\multirow{2}{*}{Dell 6420} & 88MB & \quad Video & 18.75MB & 50 & 100\% & 50 & 40\% &  50 & 6\% \\
 & 109MB & SCSI Generic & 31MB & 50 & 100\% & 50 & 54\% & 50 & 14\% \\ 
\hline
\multirow{2}{*}{Lenovo T420} & 147MB & \quad Video & 18.75MB & 50 & 100\% & 50 &  55\% & 50 & 15\% \\
 & 168MB & SCSI Generic & 31MB & 50 & 100\% & 50 & 70\% & 50 & 25\%  \\ 
 \hline
\end{tabular}
\caption{{For every attack run, the memory ambush technique always succeeds in positioning the video buffers next to the page tables (its success rate is 100\%). For each machine, a larger size of exploitable buffer indicates a higher success rate for occurrences of bit flips and exploitable bit flips, i.e., gaining the kernel privilege. (Success Rate = S.R.)}}
\label{tab:macro}
\end{table*}

\begin{table*}
\renewcommand\arraystretch{1.8}
\centering
\begin{tabular}{cccccc}
\hline
\textbf{Machine Type} & 
\textbf{Success Run} &
\textbf{Exploitable Buffer} &
\textbf{Time for Algorithm $1$} &
\textbf{Time for Algorithm $2$} & \textbf{Exploit Time} \\
\hline
 \multirow{6}{*}{Dell 6420} &
\multirow{2}{*}{Worst Case} & Video & 1\emph{sec} & 442\emph{sec} & 8\emph{min} \\
 &  & SCSI Generic & 1\emph{sec} & 3688\emph{sec} & 62\emph{min} \\
 \cline{2-6}
 & \multirow{2}{*}{Best Case} &  Video & 1\emph{sec} & 59\emph{sec} & 1\emph{min}  \\
 &  & SCSI Generic & 1\emph{sec} & 52\emph{sec} & 1\emph{min} \\
 \cline{2-6}
 & \multirow{2}{*}{Averaged} & Video & 1\emph{sec} & 191\emph{sec} & $4$\emph{min} \\
 & & SCSI Generic & 1\emph{sec} & 738\emph{sec} & 13\emph{min} \\
\hline
\multirow{6}{*}{Lenovo T420} &
 \multirow{2}{*}{Worst Case} & Video & 2\emph{sec} & 1686\emph{sec} & 29\emph{min} \\
 &  & SCSI Generic & 3\emph{sec} & 942\emph{sec} & 16\emph{min} \\
 \cline{2-6}
  & \multirow{2}{*}{Best Case} &  Video & 2\emph{sec} & 18\emph{sec} & 0.3\emph{min}  \\
 &  & SCSI Generic & 3\emph{sec} & 3\emph{sec} & 0.1\emph{min} \\
 \cline{2-6}
 & \multirow{2}{*}{Averaged} & Video & 2\emph{sec} & 580\emph{sec} & $10$\emph{min} \\
 & & SCSI Generic & 3\emph{sec}  & 270\emph{sec} & 6\emph{min} \\
\hline
\end{tabular}
\caption{{In the best case, the successful exploit of kernel privilege escalation can be done within 1 minute for the Dell and 0.1 minute for the Lenovo, respectively.}}
\label{tab:averaged}
\end{table*}

\mypara{Memory access latency distribution}
the technique of efficient hammering is highly dependent on the distribution of the memory access latency. The distribution is expected to easily distinguish DRSB from non-DRSB; otherwise it will introduce false positives, significantly reducing the efficiency of our hammering technique. 
To measure the latency, we randomly select $1,000$ pairs of page-aligned virtual addresses that are DRSB and non-DRSB, respectively. We can easily tell whether a pair of address is DRSB or not by using the Linux {\tt pagemap} and the memory module to address mapping on the Intel Sandy Bridge platform. Note that this information is only used to measure the timing channel. Our attacks do not use it directly. For each pair of addresses, we first perform read-access to them, call {\tt clflush} to flush the cpu cache lines, and then execute a memory barrier (\texttt{mfence}) to ensure that the flush operation has finished. By doing so, subsequent accesses to the addresses will be fulfilled directly from the memory (instead of the CPU cache). We repeat these steps for $5000$ times and use the \texttt{rdtscp} instruction to measure the total time used by the loop. 
The distribution of the access latency for both machines is shown in Fig.~\ref{fig:dell} and Fig.~\ref{fig:lenovo}, respectively. 
Clearly, most pairs in DRSB have a higher latency than most pairs in non-DRSB.
Based on the latency, we can perform efficient single-sided hammering and verify whether the hammering succeeds or not.

\mypara{Memory footprints}
{as shown in Algorithm~\ref{alg:ambush}, we can set the threshold through the parameter of {\tt threshold\_mem\_size}. The memory threshold refers to the size of the exploitable (double-owned) buffer, the in-memory \emph{tmp} file and the page-table pages. A lower {\tt threshold\_mem\_size} indicates a stealthier attack.
For the Dell machine running a browser, a music player and a mail client, the minimum threshold size in the setting of video buffer can be as low as 88MB, i.e., the buffer size is 18MB (18.75$\approx$18) and the \emph{tmp} file size is 2MB. The page-table size has two parts: one part is 56MB, referring to the free \emph{small blocks} and the other part is 12MB, which is size of the free DRAM rows neighboring the video-buffer rows.  
For the Lenovo machine running another typical case (i.e., a browser, a music player, a mail client and office suite), the free small block size is 115MB; thus its memory threshold size in the setting of SCSI Generic (sg) buffer is 168MB.
As such, we launch the exploit for 50 runs in each setting. The results are shown in Table~\ref{tab:macro}. 
Our exploit can succeed on both machines running different workloads, which means that the size of free small block has little impact on our attacks, but clearly and positively correlated to the occupied memory.}

\mypara{Success Rate}
{as shown in Table~\ref{tab:macro}, the \lowerambush technique in every run is effective in positioning the video buffers adjacent to the page tables, indicating that our technique essentially requires only a few memory footprints. 
Note that we can verify the adjacency by accessing a kernel module and the {\tt pagemap}. The module is developed to walk the created page tables and then return the physical addresses of the last-level page tables (PTE). The {\tt pagemap} provides the physical addresses of the video buffers. By doing so, we can obtain their DRAM layout and thus confirm that some of the video-buffer pages are neighboring the page-table pages within the same bank.} 

{For each machine, exploiting the sg buffer has a higher success rate of causing (exploitable) bit flips compared to that of video buffer. This is possibly because that the sg buffer has a larger size, thus having a higher chance to neighbor a victim row hosting page-table pages.}

{Also, the success rate of flippable bits is much higher than that of exploitable bits (i.e., a successful attack). Take the Dell machine with a video buffer as an example, within 20 out of 50 runs, the bits have been flipped in the page tables, and 3 out of 20 runs have found the exploitable bits and thus succeed, implying that an exploitable bit will occur within almost 7 occurrences of flippable bit. 
This also gives an empirical indication on other vulnerable machines. For example, the average time when the first flippable bit occurs on the Sandy Bridge i5-2500 (4GB DDR3) is 6 milliseconds~\cite{xiao2016one}, meaning that the machine needs 40 milliseconds on average (i.e., almost 7 occurrences of flippable bit) to observe an exploitable-bit occurrence.}

{Last, we conduct a double-sided rowhammer test based on a tool published by Seaborn et al.~\footnote{
https://github.com/google/rowhammer-test} on the two mentioned machines. Both machines are highly susceptible to bit flips. We run the tool on each machine for 24 hours. 2836 bit flips have been observed on the Dell while 3215 bit flips occurred on the Lenovo, indicating that the Dell is less vulnerable. This also explains why the success rates on the Dell are no greater than that of the Lenovo.}

\mypara{Exploit efficiency}
{as shown in Table~\ref{tab:averaged}, we measure the time cost that algorithms $1$ and $2$ took, respectively. Based on that, the exploit time that each successful run spends can be calculated accordingly.}

{For the Dell exploiting either video or SCSI Generic buffer, both exploits can
be done within 1 minute in the best case. For the Lenovo, the exploit in the best case can be reduced to 0.3 minute for video and 0.1 minute for SCSI Generic.}

We also select the video buffer and a memory threshold of $88$MB on the Dell to experiment with the traditional single-sided rowhammering~\cite{seaborn2015exploiting}. Its success rate is a little bit higher (about $8$\%), since it hammers different rows within same bank exhaustively. However, its average execution time is up to $72$ hours. Our attack is therefore much more efficient: we can complete the attack roughly about $4$ minutes, increasing the efficiency by $1080$ times compared to the traditional one.  

\eat{
Table~\ref{tab:comp} compares our exploit to some previous rowhammer-based attacks in terms of the memory impact. 
As shown in the table, both memory spray~\cite{seaborn2015exploiting} and memory groom~\cite{van2016drammer} require exhausting the system memory, while the memory waylay ~\cite{gruss2017another} depletes the page cache. As such, the kernel can fairly easily spot these attacks by checking the anomaly in the memory consumption. Compared to them, our \lowerambush technique can be customized with a memory threshold that is as low as 128MB and still remain effective (i.e., succeed in about 30 minutes). As such, our technique is much more stealthy than the existing techniques. 
 
\begin{table}
\footnotesize
\centering
\begin{tabular}{cccc}
\hline
{\textbf{Techniques}} & {\textbf{System Mem}} & {\textbf{Per-user Mem}} & {\textbf{Page Cache}} \\  \hline
{Memory Spray~\cite{seaborn2015exploiting}} & {Exhausted} & {Low} & {Low}\\
{Memory Groom~\cite{van2016drammer}} & {Exhausted} & {Exhausted} & {Low} \\
{Memory Waylay~\cite{gruss2017another}} & {Low} & {Low} & {Exhausted} \\
{\textbf{\upperambush}} & {Low} & {Low} & {Low} \\
\hline
\end{tabular}
\caption{A comparison of rowhammer-based techniques in terms of memory usage. Per-user memory is the amount of memory used by the process as accounted by the kernel.}
\label{tab:comp}
\end{table}
}



\eat{
\begin{table}
\footnotesize
\centering
\caption{The time cost and success rate of our attack in different memory size.}
\begin{tabular}{cccc}
\hline
\multirow{2}{*}{\textbf{Required Mem}} & \multicolumn{2}{c}{\textbf{Execution Time}} & \multirow{2}{*}{\textbf{Success Rate}} \\ 
 & Latency-Based & Random-based & \\ \hline
{$(256+18)$\emph{MB}} & {$0.31$\emph{h}} &  & {$6\%$} \\
{$(768+18)$\emph{MB}} & {$?$} & & {$?\%$} \\
{$(2048+18)$\emph{MB}} & {$?$} & {$72.33$\emph{h}} & {$?\%$} \\
\hline
\end{tabular}
\label{tab:usage}
\end{table}

\begin{table}
\footnotesize
\centering
\caption{}
\begin{tabular}{ccc}
\hline
{\textbf{Required Mem}} & {\textbf{Exploit Process}} & {\textbf{Required Page Cache}} \\  \hline
{$(256+18)$\emph{MB}} & {$2$\emph{MB}} & {$49$\emph{MB}} \\
{$(768+18)$\emph{MB}} & {$8$\emph{MB}}  & {$58$\emph{MB}} \\
{$(2048+18)$\emph{MB}} & {$16$\emph{MB}} & {$69$\emph{MB}} \\
\hline
\end{tabular}
\label{tab:impact}
\end{table}

\begin{table*}
\footnotesize
\centering
\caption{As for the introduced memory impact, the required memory including the exploit process goes up rapidly while the require page cache has a slight growth, indicating that the exploit has a major effect on system memory rather than page cache.}
\begin{tabular}{ccccccc}
\hline
\multirow{2}{*}{\textbf{Required Memory}} & \multirow{2}{*}{\textbf{Exploit Process}} & \multirow{2}{*}{\textbf{Required Page Cache}} & \multicolumn{2}{c}{\textbf{Latency-based Rowhammer}} & \multicolumn{2}{c}{\textbf{Random-base Rowhammer}} \\
 &  &  & Execution Time & Success Rate & Execution Time & Success Rate \\ \hline
\multirow{1}{*}{$(256+18)$\emph{MB}} & {$2$\emph{MB}} & {$49$\emph{MB}} & $0.3$\emph{h} & $6\%$ & $103$\emph{h} & $6\%$ \\
\hline
\multirow{1}{*}{$(768+18)$\emph{MB}} & {$8$\emph{MB}} & {$58$\emph{MB}} & $0.2$\emph{h} & $10\%$ & $61$\emph{h} & $14\%$ \\
\hline
\multirow{1}{*}{$(2048+18)$\emph{MB}} & {$16$\emph{MB}} & {$69$\emph{MB}} & $0.2$\emph{h} & $12\%$ & $72$\emph{h} & $6\%$ \\
\hline
\end{tabular}
\label{tab:cases}
\end{table*}

}





\section{Mitigation}\label{sec:mitigation}
As we have demonstrated so far, CATT's static kernel and user partition is ineffective in the face of double-owned memory. Allocating double-owned memory either from the kernel memory or the user memory does not seem to be secure: the former exposes the kernel to rowhammer attacks, while the latter has the same bad effect (e.g., exposing the device drivers and security-sensitive modules to the same attacks). Our exploit has demonstrated that the former is not secure. The latter is likely insecure as well. For instance, device drivers are notoriously vulnerable and recent years' research has shown that hardware devices, even the CPU, are not immune from vulnerabilities~\cite{LXFI2011software}. 

To temporarily work against our current video/sg-buffer-based attacks, we can borrow the idea from CATT by isolating the double-owned memory for the video/sg driver. Specifically, we allocate the video/sg buffer using physically continuous pages and leave one guard row on each side of the buffer. By doing so, hammering the video/sg buffer will only affect the buffer itself and it is also protected from hammering the security-sensitive objects. This wastes two rows per device driver buffer, i.e., 16KB memory on our test platform. Given that modern computers often have more than 8GB of memory and a computer has limited number of devices that require double-owned buffer, the memory waste does not seem to be a problem at all. 

However, this CATT's idea cannot be applied to isolate all the shared/mmapped buffers used by the kernel modules shown in Fig.~\ref{fig:dist}, as this would result in numerous domains as well as memory fragments. Furthermore, Fig.~\ref{fig:mmap} shows that the number of mapped buffers is still growing rapidly. It is impractical for CATT to implement too many domains. 
Alternatively, disabling the \texttt{mmap} feature is straightforward but not realistic. It requires large engineering efforts to restructure all the affected kernel modules (i.e., replace the mmaped buffer with two buffers and make extra copies from one to the other). Inevitably, this would lead to high performance loss.  
As such, we intend to explore a practical defense against our exploit in our future work. 


%
%

\section{Discussion}\label{sec:dis}
In this section, we will discuss possible improvements to our system. 

{\mypara{Eliminate fresh small blocks}
after available small blocks are depleted in step \textbf{(B)} and before we proceed to step \textbf{(C)} in Figure~\ref{fig:ambush}, it is likely for target blocks or large blocks to be split for other user processes or the kernel, thus introducing new small blocks. Also some allocated small blocks might be freed by other processes and thus become available right before step \textbf{(C)}. In such a case, we will increase the memory threshold a bit so that we can consume the dynamically created small blocks before we start step \textbf{(C)}.}

\mypara{Obtain the virtual-to-physical address mapping}
by leveraging the prefetch side channel~\cite{gruss2016prefetch}, an adversary can obtain the virtual-to-physical address mapping without {\tt pagemap}, making it possible again to perform the double-sided hammering. However, a recent kernel patch called KAISER~\cite{gruss2017kaslr} (also known as kernel page table isolation) protects against the channel and has been widely applied in recent Linux kernel versions. Note that the memory waylaying technique~\cite{gruss2017another} that relies on the side channel will no longer be applicable in such Linux kernels.

\mypara{Make the attack stealthier} 
like other rowhammer attackers, our exploit has specific instructions or abnormal memory access patterns that can be detected by static or dynamic analysis tools. Also, our exploit has high cache miss rates, which can be observed by monitoring CPU performance counters.
For example, MASCAT~\cite{irazoqui2016mascat} performs a static code analysis of a target application to detect state-of-the-art DRAM access attacks, including the rowhammer attacks. 
ANVIL~\cite{aweke2016anvil} uses the hardware performance counters to monitor the miss rate of the last-level CPU cache. Whenever the rate is high enough to conduct a rowhammer attack, ANVIL will be triggered to further analyze the process for malicious behaviors. Further, it can discover this unusual access pattern and use heuristics to identify a potential rowhammer attack. 
 
Such countermeasures can be bypassed by applying both one-location hammering and Intel Software Guard Extension (SGX)~\cite{costan2016intel}. Although the one-location hammering induces less bit flips compared to the other two rowhammer methods, this technique just keeps opening and closing one row, making itself stealthy to bypass ANVIL.
Intel SGX is a hardware extension in Intel CPUs to securely run trusted code in an untrusted system. We can hide our rowhammer exploit inside an SGX enclave, where the exploit code cannot be analyzed by the other software because any external access to the enclave is denied. Features like performance counters and debug registers also cannot be used to monitor the enclave activities~\cite{costan2016intel}. In particular, Schwarz et al.~\cite{schwarz2017malware} have confirmed that performance counters will not record the CPU cache hit or miss data inside the enclave. 





\section{Related Work}\label{sec:related}
In this section, we compare our system to the existing rowhammer attacks and discuss the related defenses. 

\subsection{Rowhammer Attacks}

\eat{
\emph{\textbf{Process Isolation.}} Seaborn et al.~\cite{seaborn2015exploiting} use x86 \emph{clflush} instruction and memory spraying to escape the Native Client sandbox within a browser. Similarly, Gruss et al.~\cite{gruss2016rowhammer} crafts special memory access pattern to escape Linux scripting environment and Bosman et al.~\cite{bosman2016dedup} reply on \emph{page-deduplication}~\cite{pagededuplication} to compromise Windows-based JavaScript isolation. 
    Bhattacharya et al.~\cite{bhattacharya2016curious} utilize \emph{pagemap}~\cite{shutemovpagemap} to cause bit-flips in a running exponentiation algorithm and thus recover a secret key. Gruss et al.~\cite{gruss2017another} exploit \emph{page cache}~\cite{pagecache} to flip opcodes in a \emph{setuid} running program and thus gain root privilege. 
    \emph{\textbf{Kernel Isolation.}}  
    These attacks~\cite{seaborn2015exploiting, qiao2016new, aweke2016anvil, van2016drammer} modify page tables entries (PTEs) to gain kernel privilege through different attack vectors. Both Seaborn and Qiao~\cite{seaborn2015exploiting, qiao2016new} utilize specific x86 CPU instructions, i.e., \emph{clflush} and \emph{movnti} respectively while Aweke et al.~\cite{aweke2016anvil} construct a specific memory-access pattern to avoid reading from the caches. Drammer~\cite{van2016drammer} aims at the ARM architecture.  
    \emph{\textbf{Virtual Machine (VM) Isolation.}} A VM owner can completely control a target VM that is within the same host by the way of page deduplication~\cite{razavi2016flip} or page-table manipulation~\cite{xiao2016one}.
    \emph{\textbf{Hypervisor Isolation.}} A VM owner can also gain the privilege of hypervisor by inducing bit-flips in PTEs~\cite{xiao2016one}.

Based on their code template that use x86 \emph{clflush} instruction to explicitly flush cache, Seaborn et al.~\cite{seaborn2015exploiting} continued the research and demonstrated the NaCI sandbox escape and kernel-privilege escalation. In the first attack, they filled available memory with authentication code and randomly picked up pairs of addresses to rowhammer the code. In the second attack, with the knowledge of pagemap and DRAM address mapping~\cite{seaborndram}, they . By doing so, efficient rowhammering can be conducted.

Bhattacharya et al.~\cite{bhattacharya2016curious}  \emph{prime+probe}~\cite{percival2005cache, osvik2006cache} and double-sided rowhammer attack to induce bit flips in memory where RSA keys are stored. 

??? Gruss et al.~\cite{gruss2016rowhammer} memory spraying launched memory 
demonstrated that rowhammer can be launched from JavaScript. Specifically, they were able to launch an attack against the page tables in a recent Firefox version.

??? Bosman et al.~\cite{bosman2016dedup} extended this work by exploiting the page deduplication feature. windows 10 also launched from JavaScript, gain arbitrary read/write access within a browser. 

??? Razvi et al.~\cite{razavi2016flip} also exploits page deduplication and rowhammer attack to attack a co-located VM and obtain its private keys.

In a paravirtual cloud, Xiao et al.~\cite{xiao2016one} had the access to \emph{pagemap} reverse-engineered DRAM address mapping, sprayed page tables, inside which bit flips were induced. 

??? Van et al.~\cite{van2016drammer} utilized the memory grooming technique to surgically place page tables onto vulnerable DRAM rows by . DMA buffer without the \emph{pagemap}. 

The known approaches to
solve this challenge are spraying, i.e., filling the entire memory
with copies of the page, or grooming, i.e., allocating the target
page in exactly the right moment

Cross-VM Row-Hammer Attacks~\cite{xiao2016one} are demonstrated to break Xen memory isolation enforced by paravirtualized Xen and gain 

cross-VM settings,
in which a malicious VM exploits bit flips induced
by row hammer attacks to crack memory isolation enforced
by virtualization. 
exhaustively hammer a large fraction of physical
memory from a guest VM (to collect exploitable
vulnerable bits), and innovative approaches to break
Xen paravirtualized memory isolation (to access arbitrary
physical memory of the shared machine).

\cite{razavi2016flip,qiao2016new,bosman2016dedup,gruss2016rowhammer,brasser17can,seaborn2015exploiting}
}

\eat{
\zhi{it has three other disadvantages: 1. it requires CPU to feature Intel SGX~\cite{costan2016intel, mckeen2013innovative} while our attack does not need any special hardware features. 2. there are false negatives when it is determining whether a target page is at the expected physical memory location. (address-translation) 3. a single process exhausts page cache pages, which is easily detected by the OS administrator.}
}

We first review how rowhammer attacks achieve the different requirements, specifically, how the CPU cache is flushed, how the row buffer is cleared, and how the aggressor and victim rows are placed. 

\mypara{Bypass CPU cache}
since frequent and direct memory access is a prerequisite for hammering,  a simple solution is to use the  \texttt{clflush} instruction for explicit CPU cache flush~\cite{kim2014flipping, seaborn2015exploiting}. 
This instruction can flush cache entries related to a specific virtual address, and thus subsequent read to the address will be served directly from the memory. \texttt{clflush}  is included in the instruction set for a process to fetch the updated data from the memory instead of the obsolete cached ones. It can be executed by an unprivileged process. It has been proposed to prohibit user processes to execute the instruction as a defense against rowhammer attacks~\cite{seaborn2015exploiting}. 
However, Qiao et al.~\cite{qiao2016new} reported that commonly used x86 instructions such as \texttt{movnti} and \texttt{movntdqa} actually bypass the CPU cache and access the memory directly. 
Moreover, carefully crafted memory-access patterns~\cite{aweke2016anvil,bhattacharya2016curious,gruss2016rowhammer,bosman2016dedup,frigo2018grand} can cause cache conflict and effectively evict the cache of the target address. This approach is particularly useful for the scripting environments where cache-related instructions are not directly available. 
At last, DMA-based memory is uncached by the CPU cache and thus it is exploited by multiple attacks~\cite{tatar2018throwhammer,lipp2018nethammer,van2018guardion,van2016drammer} to directly reach DRAM.

\mypara{Clear row buffer}  
besides flushing the cache, rowhammer attacks also need to clear the row buffer in order to keep ``opening'' a row. Different rowhammer attacks have achieved this goal with various techniques.

Double-sided hammering performs alternate reads on different rows in the same bank. Therefore, it requires the virtual-to-physical and physical-to-hardware mappings in order to position the aggressor and victim rows. The \texttt{pagemap} provides complete information of the first mapping, but it is not accessible to the unprivileged process now. Although huge page on x86~\cite{seaborn2015exploiting} and the DMA buffers on the ARM architecture~\cite{van2016drammer} only give the partial information about the virtual-to-physical address mapping, they ensure that two virtually-continuous addresses are also physically continuous. They can also be used by rowhammer attacks. Alternatively, Glitch~\cite{frigo2018grand} relies on precise GPU-based timers to detect physically-contiguous memory. For the second mapping, AMD provides the details in their manual, and the mapping for various Intel CPUs has been reverse engineered~\cite{xiao2016one,pessl2016drama}. 

In contrast, single-sided hammering~\cite{seaborn2015exploiting} and one-location hammering~\cite{gruss2017another} do not need both mappings to clear the row buffer. To this end, single-sided hammering randomly selects multiple virtual addresses to access and it is likely that these addresses are in different rows within the same bank, thus clearing the row buffer.
{For one-location hammering, it keeps opening one randomly chose row and leverages the advanced DRAM controller to close the row. Both single-sided and one-location hammering hammer a row in a less frequent way than that of double-sided hammering, thus they are less efficient in inducing bit flips.}

\mypara{Place target objects}
the last requirement of rowhammer attacks is to manipulate the security domain into placing a security-sensitive object in a vulnerable row. This can be achieved through page-table spraying~\cite{seaborn2015exploiting,gruss2016rowhammer}, page deduplication~\cite{bosman2016dedup,razavi2016flip}, Phys Feng Shui~\cite{van2016drammer}, RAMpage~\cite{van2018guardion} and Throwhammer~\cite{tatar2018throwhammer}. However, they all require exhausting the memory in order to place the target page in the vulnerable row. 
Instead of depleting the system memory, memory waylay~\cite{gruss2017another} exhausts the page cache to influence the physical location of a target page. Without exhausting either memory or page cache, Glitch~\cite{frigo2018grand} and our memory-ambush technique can satisfy this requirement with a much constrained amount of memory but Glitch only gains the browser privilege.

\subsection{Rowhammer Defenses}
Both hardware and software defense against rowhammer attacks have been proposed. 
Hardware defenses can be based on the firmware or new hardware designs. For example, computer manufacturers, such as HP~\cite{HP}, Lenovo~\cite{LENOVO} and Apple~\cite{apple}, propose to double the refresh rate of DRAM from 64ms to 32ms. This slightly raises the bar for the attack but has been proven to be ineffective~\cite{aweke2016anvil}.
Intel suggests to use Error Correcting Code (ECC) memory to catch and correct single-bit errors on-the-fly, thus alleviating bit flips by rowhammer attacks~\cite{intelecc}. Typically, ECC can correct single-bit errors and detect double-bit errors (e.g., SECDED). However, ECC cannot prevent multiple bit errors and normally is only available on the server systems.
Probabilistic adjacent row activation (PARA)~\cite{kim2014flipping} activates/refreshes adjacent rows with a high probability when the aggressor rows are hammered many times. This could be effective but needs changes of the memory controller. 
For future DRAM architectures, new DDR4 modules~\cite{DDR4} and LPDDR4 specification~\cite{lpDDR4} propose a targeted row refresh (TRR) capability to mitigate rowhammer attacks. However, Nethammer~\cite{lipp2018nethammer} presented a network-based rowhammer attack in the face of TRR, causing kernel crashes.  

Many software-based defenses have also been proposed. Some defenses aim to preventing attacks from misusing specific system features.  For example, researchers and developers have worked to prevent the {\tt pagemap}~\cite{shutemovpagemap,seaborn2015exploiting}, page deduplication~\cite{pagededuplication}, specific x86 CPU instructions~\cite{seaborn2015exploiting}, GPU timers~\cite{glitch}, ION contiguous heap~\cite{ION}  and memory/page-cache exhaustion~\cite{gruss2016rowhammer,van2016drammer, gruss2017another} from being abused by unprivileged attackers. 
ANVIL~\cite{aweke2016anvil} is the first system to detect rowhammer behaviors using the Intel hardware performance counters~\cite{intelOp}. However, ANVIL incurs a high performance overhead in its worst case and has false positives~\cite{brasser17can}. 
Brasser et al.~\cite{brasser2016can} patch an open-source bootloader and disable the vulnerable memory modules. Although this approach effectively eliminates all the rowhammer
vulnerabilities for legacy systems, it is not practical when most memory is susceptible to rowhammer and this method is not compatible with Windows. 

Inspired by the CATT concept, there
are other three software-only techniques achieving orthogonal instances of the physical domain isolation, that is, GuardION~\cite{van2018guardion} physically isolated DMA memory. ALIS~\cite{tatar2018throwhammer} presented physical memory isolation for network relevant memory while RIP-RH~\cite{bock2019rip} enforced physical memory isolation for target user processes. We believe that such defense techniques based on the CATT concept are also not
secure if their deployment in real-world don't carefully consider the performance design of modern OSes (i.e., they can be identified to have a similar memory-ownership issue).

\section{Conclusion}\label{sec:conclusion}
In this paper, we presented a novel practical exploit, which could effectively defeat the phyical kernel isolation and gain the root and kernel privileges. Our exploit does not need to exhaust the page cache or the system memory. 
In addition, it does not rely on the virtual-to-physical address mapping information. 
To achieve these unique features, we proposed the \lowerambush technique, which leverages the inherent Linux memory management features, to make our exploit stealthy. 
We improved the single-sided hammering by utilizing the timing channel caused by the row buffer. 
We have implemented two proof-of-concept attacks on the Linux platform. The experiment results show that our exploit can complete in less than $1$ minute and require memory as low as $88$\emph{MB}.

\ifCLASSOPTIONcaptionsoff
  \newpage
\fi



%



{\footnotesize \bibliographystyle{IEEEtranS}
\bibliography{main}}

\begin{thebibliography}{10}
\providecommand{\url}[1]{#1}
\csname url@samestyle\endcsname
\providecommand{\newblock}{\relax}
\providecommand{\bibinfo}[2]{#2}
\providecommand{\BIBentrySTDinterwordspacing}{\spaceskip=0pt\relax}
\providecommand{\BIBentryALTinterwordstretchfactor}{4}
\providecommand{\BIBentryALTinterwordspacing}{\spaceskip=\fontdimen2\font plus
\BIBentryALTinterwordstretchfactor\fontdimen3\font minus
  \fontdimen4\font\relax}
\providecommand{\BIBforeignlanguage}[2]{{%
\expandafter\ifx\csname l@#1\endcsname\relax
\typeout{** WARNING: IEEEtranS.bst: No hyphenation pattern has been}%
\typeout{** loaded for the language `#1'. Using the pattern for}%
\typeout{** the default language instead.}%
\else
\language=\csname l@#1\endcsname
\fi
#2}}
\providecommand{\BIBdecl}{\relax}
\BIBdecl

\bibitem{apple}
{Apple, Inc.}, ``About the security content of mac efi security update
  2015-001,'' \url{https://support.apple.com/en-au/HT204934}, Aug. 2015.

\bibitem{aweke2016anvil}
Z.~B. Aweke, S.~F. Yitbarek, R.~Qiao, R.~Das, M.~Hicks, Y.~Oren, and T.~Austin,
  ``Anvil: Software-based protection against next-generation rowhammer
  attacks,'' \emph{ACM SIGPLAN Notices}, vol.~51, no.~4, pp. 743--755, 2016.

\bibitem{bhattacharya2016curious}
S.~Bhattacharya and D.~Mukhopadhyay, ``Curious case of rowhammer: flipping
  secret exponent bits using timing analysis,'' in \emph{International
  Conference on Cryptographic Hardware and Embedded Systems}.\hskip 1em plus
  0.5em minus 0.4em\relax Springer, 2016, pp. 602--624.

\bibitem{bock2019rip}
C.~Bock, F.~Brasser, D.~Gens, C.~Liebchen, and A.-R. Sadeghi, ``Rip-rh:
  Preventing rowhammer-based inter-process attacks,'' in \emph{Proceedings of
  the 2019 ACM Asia Conference on Computer and Communications Security}.\hskip
  1em plus 0.5em minus 0.4em\relax ACM, 2019, pp. 561--572.

\bibitem{bosman2016dedup}
E.~Bosman, K.~Razavi, H.~Bos, and C.~Giuffrida, ``Dedup est machina: Memory
  deduplication as an advanced exploitation vector,'' in \emph{Security and
  Privacy, 2016 IEEE Symposium on}.\hskip 1em plus 0.5em minus 0.4em\relax
  IEEE, 2016, pp. 987--1004.

\bibitem{brasser2016can}
F.~Brasser, L.~Davi, D.~Gens, C.~Liebchen, and A.-R. Sadeghi, ``Can't touch
  this: Practical and generic software-only defenses against rowhammer
  attacks,'' \emph{arXiv preprint arXiv:1611.08396}, 2016.

\bibitem{brasser17can}
------, ``Can't touch this: Software-only mitigation against rowhammer attacks
  targeting kernel memory,'' in \emph{USENIX Security Symposium}, 2017.

\bibitem{highsecure}
K.~Cook, ``kexec: add sysctl to disable kexec\_load,''
  \url{https://lwn.net/Articles/580269/}, 2014.

\bibitem{costan2016intel}
V.~Costan and S.~Devadas, ``Intel sgx explained.'' \emph{IACR Cryptology ePrint
  Archive}, vol. 2016, p.~86, 2016.

\bibitem{frigo2018grand}
P.~Frigo, C.~Giuffrida, H.~Bos, and K.~Razavi, ``Grand pwning unit:
  accelerating microarchitectural attacks with the gpu,'' in \emph{Security and
  Privacy, 2018 IEEE Symposium on}.\hskip 1em plus 0.5em minus 0.4em\relax
  IEEE, 2018.

\bibitem{ION}
{Google, Inc.}, ``Disable ion heap type system contig,'' \url{https: //
  android.googlesource.com/device/google/marlin-kernel/+/android-7.1.0_r7},
  2016.

\bibitem{glitch}
------, ``Glitch vulnerability status,''
  \url{http://www.chromium.org/chromium-os/glitch-vulnerability-status}, May
  2018.

\bibitem{gorman2004understanding}
M.~Gorman, \emph{Understanding the Linux virtual memory manager}.\hskip 1em
  plus 0.5em minus 0.4em\relax Prentice Hall Upper Saddle River, 2004.

\bibitem{gruss2017kaslr}
D.~Gruss, M.~Lipp, M.~Schwarz, R.~Fellner, C.~Maurice, and S.~Mangard, ``Kaslr
  is dead: long live kaslr,'' in \emph{International Symposium on Engineering
  Secure Software and Systems}.\hskip 1em plus 0.5em minus 0.4em\relax
  Springer, 2017, pp. 161--176.

\bibitem{gruss2017another}
D.~Gruss, M.~Lipp, M.~Schwarz, D.~Genkin, J.~Juffinger, S.~O'Connell,
  W.~Schoechl, and Y.~Yarom, ``Another flip in the wall of rowhammer
  defenses,'' \emph{arXiv preprint arXiv:1710.00551}, 2017.

\bibitem{gruss2016prefetch}
D.~Gruss, C.~Maurice, A.~Fogh, M.~Lipp, and S.~Mangard, ``Prefetch side-channel
  attacks: Bypassing smap and kernel aslr,'' in \emph{Proceedings of the 2016
  ACM SIGSAC conference on computer and communications security}.\hskip 1em
  plus 0.5em minus 0.4em\relax ACM, 2016, pp. 368--379.

\bibitem{gruss2016rowhammer}
D.~Gruss, C.~Maurice, and S.~Mangard, ``Rowhammer.js: A remote software-induced
  fault attack in javascript,'' in \emph{Detection of Intrusions and Malware,
  and Vulnerability Assessment}.\hskip 1em plus 0.5em minus 0.4em\relax
  Springer, 2016, pp. 300--321.

\bibitem{rowhammerjs}
------, ``Program for testing for the dram rowhammer problem using eviction,''
  \url{https://github.com/IAIK/rowhammerjs}, May 2017.

\bibitem{HP}
{HP, Inc.}, ``Hp moonshot component pack,''
  \url{https://support.hpe.com/hpsc/doc/public/display?docId=c04676483}, May
  2015.

\bibitem{intelOp}
{Intel, Inc.}, ``Intel 64 and {IA}-32 architectures optimization reference
  manual,'' Sep. 2014.

\bibitem{intelecc}
------, ``The role of ecc memory,''
  \url{https://www.intel.com/content/www/us/en/workstations/workstation-ecc-memory-brief.html},
  2015.

\bibitem{irazoqui2016mascat}
G.~Irazoqui, T.~Eisenbarth, and B.~Sunar, ``Mascat: Stopping microarchitectural
  attacks before execution.'' \emph{IACR Cryptology ePrint Archive}, 2016.

\bibitem{lpDDR4}
{JEDEC Solid State Technology Association.}, ``Low power double data rate 4
  (lpddr4),'' \url{https://www.jedec.org/standards-documents/docs/jesd209-4b},
  2015.

\bibitem{kim2014flipping}
Y.~Kim, R.~Daly, J.~Kim, C.~Fallin, J.~H. Lee, D.~Lee, C.~Wilkerson, K.~Lai,
  and O.~Mutlu, ``Flipping bits in memory without accessing them: An
  experimental study of dram disturbance errors,'' in \emph{ACM SIGARCH
  Computer Architecture News}, vol.~42, no.~3.\hskip 1em plus 0.5em minus
  0.4em\relax IEEE Press, 2014, pp. 361--372.

\bibitem{LENOVO}
{LENOVO, Inc.}, ``Row hammer privilege escalation lenovo security advisory:
  Len-2015-009,''
  \url{https://support.lenovo.com/au/en/product_security/row_hammer}, Aug.
  2015.

\bibitem{v4l2}
Linux, ``Video for linux api,''
  \url{https://www.kernel.org/doc/html/v4.12/media/uapi/v4l/v4l2.html}, July
  2016.

\bibitem{lipp2018nethammer}
M.~Lipp, M.~T. Aga, M.~Schwarz, D.~Gruss, C.~Maurice, L.~Raab, and L.~Lamster,
  ``Nethammer: Inducing rowhammer faults through network requests,''
  \emph{arXiv preprint arXiv:1805.04956}, 2018.

\bibitem{LXFI2011software}
Y.~Mao, H.~Chen, D.~Zhou, X.~Wang, N.~Zeldovich, and M.~F. Kaashoek, ``Software
  fault isolation with api integrity and multi-principal modules,'' in
  \emph{Proceedings of the Twenty-Third ACM Symposium on Operating Systems
  Principles}.\hskip 1em plus 0.5em minus 0.4em\relax ACM, 2011, pp. 115--128.

\bibitem{DDR4}
{Micron, Inc.}, ``Ddr4 sdram mt40a2g4, mt40a1g8, mt40a512m16 data sheet,''
  \url{https://www.micron.com/products/dram/ddr4-sdram/}, 2015.

\bibitem{pagededuplication}
Microsoft, ``Cache and memory manager improvements,''
  \url{https://docs.microsoft.com/en-us/windows-server/administration/performance-tuning/subsystem/cache-memory-management/improvements-in-windows-server},
  2017.

\bibitem{moscibroda2007memory}
T.~Moscibroda and O.~Mutlu, ``Memory performance attacks: Denial of memory
  service in multi-core systems,'' in \emph{USENIX Security Symposium}, 2007.

\bibitem{pessl2016drama}
P.~Pessl, D.~Gruss, C.~Maurice, M.~Schwarz, and S.~Mangard, ``Drama: Exploiting
  dram addressing for cross-cpu attacks,'' in \emph{USENIX Security Symposium},
  2016, pp. 565--581.

\bibitem{qiao2016new}
R.~Qiao and M.~Seaborn, ``A new approach for rowhammer attacks,'' in
  \emph{Hardware Oriented Security and Trust (HOST), 2016 IEEE International
  Symposium on}.\hskip 1em plus 0.5em minus 0.4em\relax IEEE, 2016, pp.
  161--166.

\bibitem{razavi2016flip}
K.~Razavi, B.~Gras, E.~Bosman, B.~Preneel, C.~Giuffrida, and H.~Bos, ``Flip
  feng shui: Hammering a needle in the software stack,'' in \emph{USENIX
  Security Symposium}, 2016, pp. 1--18.

\bibitem{schwarz2017malware}
M.~Schwarz, S.~Weiser, D.~Gruss, C.~Maurice, and S.~Mangard, ``Malware guard
  extension: Using sgx to conceal cache attacks,'' \emph{arXiv preprint
  arXiv:1702.08719}, 2017.

\bibitem{seaborn2015exploiting}
M.~Seaborn and T.~Dullien, ``Exploiting the dram rowhammer bug to gain kernel
  privileges,'' in \emph{Black Hat'15}.

\bibitem{shutemovpagemap}
K.~A. Shutemov, ``Pagemap: Do not leak physical addresses to non-privileged
  userspace,'' \url{https://lwn.net/Articles/642074/}, 2015.

\bibitem{soltesz2007container}
S.~Soltesz, H.~P{\"o}tzl, M.~E. Fiuczynski, A.~Bavier, and L.~Peterson,
  ``Container-based operating system virtualization: a scalable,
  high-performance alternative to hypervisors,'' in \emph{ACM SIGOPS Operating
  Systems Review}.\hskip 1em plus 0.5em minus 0.4em\relax ACM, 2007, pp.
  275--287.

\bibitem{tatar2018throwhammer}
A.~Tatar, R.~K. Konoth, E.~Athanasopoulos, C.~Giuffrida, H.~Bos, and K.~Razavi,
  ``Throwhammer: Rowhammer attacks over the network and defenses,'' in
  \emph{2018 USENIX Annual Technical Conference}, 2018.

\bibitem{van2016drammer}
V.~van~der Veen, Y.~Fratantonio, M.~Lindorfer, D.~Gruss, C.~Maurice, G.~Vigna,
  H.~Bos, K.~Razavi, and C.~Giuffrida, ``Drammer: Deterministic rowhammer
  attacks on mobile platforms,'' in \emph{Proceedings of the 2016 ACM SIGSAC
  Conference on Computer and Communications Security}.\hskip 1em plus 0.5em
  minus 0.4em\relax ACM, 2016, pp. 1675--1689.

\bibitem{van2018guardion}
V.~van~der Veen, M.~Lindorfer, Y.~Fratantonio, H.~P. Pillai, G.~Vigna,
  C.~Kruegel, H.~Bos, and K.~Razavi, ``Guardion: Practical mitigation of
  dma-based rowhammer attacks on arm,'' in \emph{International Conference on
  Detection of Intrusions and Malware, and Vulnerability Assessment}.\hskip 1em
  plus 0.5em minus 0.4em\relax Springer, 2018, pp. 92--113.

\bibitem{xiao2016one}
Y.~Xiao, X.~Zhang, Y.~Zhang, and R.~Teodorescu, ``One bit flips, one cloud
  flops: Cross-vm row hammer attacks and privilege escalation,'' in
  \emph{USENIX Security Symposium}, 2016, pp. 19--35.

\end{thebibliography}

%
\vspace{-1.0cm}
\begin{IEEEbiography}[{\includegraphics[width=1in,height=1.25in,clip,keepaspectratio]{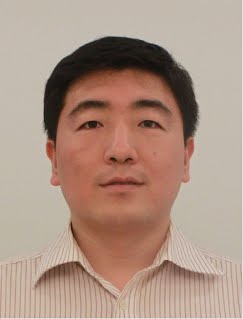}}]%
{Yueqiang Cheng} is a Staff Security Scientist at Baidu XLab America. He earned his PhD degree in School of Information Systems from Singapore Management University under the guidance of Professor Robert H. Deng and Associate Professor Xuhua Ding. 
His research interests are system security, trustworthy computing, software-only root of trust and software security.
\end{IEEEbiography}
\vspace{-1.0cm}
\begin{IEEEbiography}[{\includegraphics[width=1in,height=1.25in,clip]{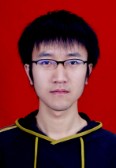}}]%
{Zhi Zhang} is a PhD student in the School of Computer Science and Engineering at the University of New South Wales. His research interests are in the areas of system security and virtualization. He received his bachelor degree from Sichuan University in 2011 and his master degree from Peking University in 2014.
\end{IEEEbiography}
\vspace{-1.0cm}
\begin{IEEEbiography}[{\includegraphics[width=1in,height=1.25in,clip]{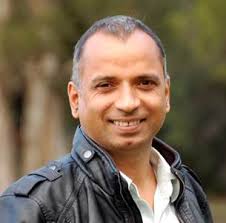}}]%
{Surya Nepal} is a Principal Research Scientist at CSIRO Data61 and leads the distributed system security research group. His main research focus has been in the area of distributed systems and social networks, with a specific focus on security, privacy and trust. He has more than 200 peer-reviewed publications to his credit. He currently serves as an associate editor in an editorial board of IEEE Transactions on Service Computing.
\end{IEEEbiography}
\vspace{-1.0cm}
\begin{IEEEbiography}[{\includegraphics[width=1in,height=1.25in,clip]{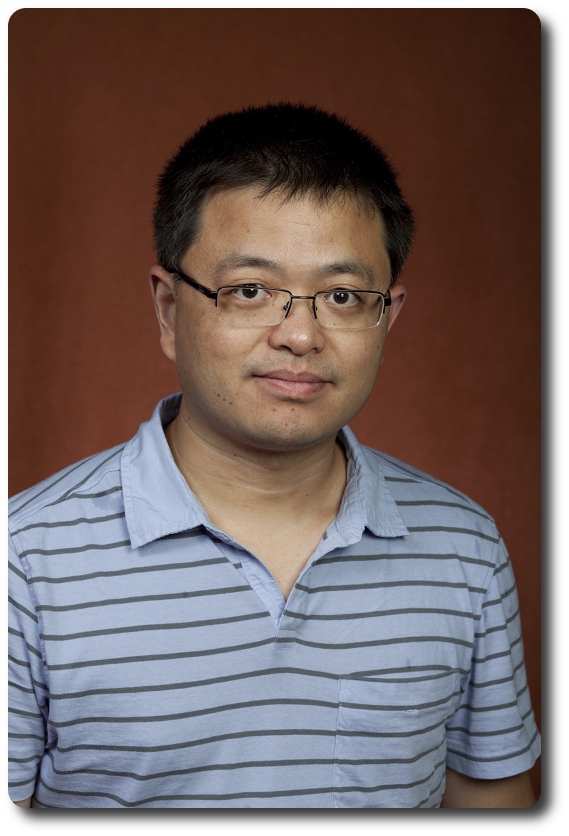}}]%
{Zhi Wang} is an Associate Professor in the Department of Computer Science at the Florida State University. He has broad research interests in security with a focus on the systems security, particularly, operating systems\/virtualization security, software security, and mobile security.
\end{IEEEbiography}






\end{document}